\documentclass[prd,aps,
twocolumn,
superscriptaddress]{revtex4}

\usepackage[dvipsnames]{xcolor}
\usepackage{graphicx}
\usepackage{amssymb}

\usepackage{amsthm}
\usepackage{amsmath}
\usepackage{txfonts}

\usepackage{dcolumn}
\usepackage{lipsum}

\usepackage{soul}
\usepackage{url}
\usepackage{epsfig}
\usepackage{bm}
\usepackage{setspace}
\usepackage{appendix}
\usepackage{lscape}

\usepackage{bbold}
\usepackage{dcolumn}
\usepackage[utf8]{inputenc}

\usepackage{natbib}
\usepackage{bm}
\usepackage{amsmath}
\usepackage{float}
\usepackage{multirow}
\usepackage{slashed}
\usepackage{xcolor}
\usepackage{physics}
\usepackage{multirow}
\usepackage{gensymb}
\usepackage{mathtools,braket}
\usepackage{subcaption}
\usepackage{lipsum}  
\usepackage{soul}
\usepackage[colorlinks=true,
            linkcolor=red,
            citecolor=blue,
            urlcolor=blue]{hyperref}
\usepackage{bm}
\usepackage{xspace}
\usepackage{cancel}
\usepackage{float}
\usepackage{multirow}
\definecolor{darkgreen}{rgb}{0,0.5,0}
\definecolor{purple}{rgb}{0.5,0,0.5}
\definecolor{nblue}{rgb}{0.0,0.0,0.50}
\definecolor{scarlet}{rgb}{1.0,0.2,0}
\definecolor{darkmagenta}{rgb}{0.55, 0.0, 0.55}
\definecolor{darkolivegreen}{rgb}{0.33, 0.42, 0.18}
\definecolor{darkcandyapplered}{rgb}{0.64, 0.0, 0.0}
\usepackage{cleveref}

\newcommand{\orcid}[1]{\href{https://orcid.org/#1}{\includegraphics[width=8pt]{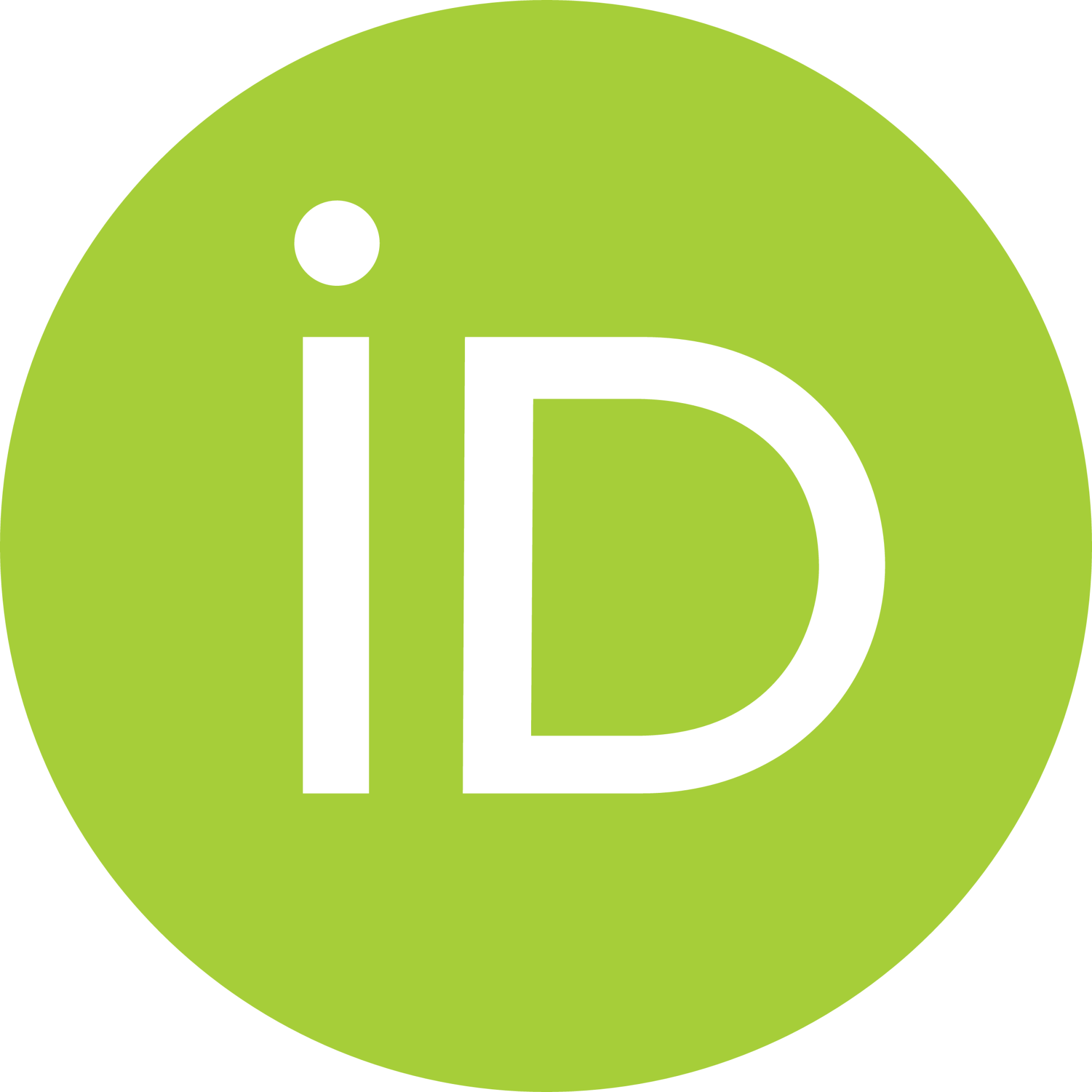}}}

\newcommand{\be}{\begin{equation}}

\newcommand{\ee}{\end{equation}}
\newcommand{\bea}{\begin{eqnarray}}
\newcommand{\eea}{\end{eqnarray}}
\newcommand{\beas}{\begin{eqnarray*}}
\newcommand{\eeas}{\end{eqnarray*}}

\begin{document}
\title{Extraction of Pion Unpolarized Quark and Gluon Generalized Parton Distributions using Deep neural-networks}

\author{Satyajit Puhan}
\email[]{puhansatyajit@gmail.com}
\affiliation{Institute of Physics, Academia Sinica, Taipei 11529, Taiwan}

\author{Shubham Sharma}
\email[]{s.sharma.hep@gmail.com}
\affiliation{Moscow Institute of Physics and Technology, Dolgoprudny 141700, Russia}

\author{Narinder Kumar}
\email[]{narinderhep@gmail.com}
\affiliation{Computational Theoretical High Energy Physics Lab, Department of Physics, Doaba College, Jalandhar 144004, India}

\begin{abstract}
We present a deep neural-network (DNN) extraction  of the pion unpolarized quark and gluon generalized parton distributions (GPDs) using the corresponding parton distribution functions (PDFs) from the JAM21 and xFitter analysis, together with experimental measurements of the pion electromagnetic form factor (EMFF) and lattice quantum chromodynamics (QCD) results. The GPDs are parameterized using a physics-informed neural-network (PINN) that incorporates the known PDF behavior, an exponential momentum-transfer dependence, and a trainable neural network (NN) component. The network parameters are determined by minimizing a $\chi^2$-based loss function. For the valence-quark GPDs, the loss function includes contributions from the EMFF, squared EMFF, charge-normalization constraints, and regularization terms. For the gluon GPDs, it incorporates constraints from the gluon gravitational form factors together with regularization. This framework enables a flexible, nonparametric extraction while preserving the essential theoretical and phenomenological constraints. By employing the full ensemble of available PDF replicas, we quantify the uncertainties of the extracted GPDs over a broad kinematic range in the longitudinal momentum fraction and momentum transfer, with the uncertainty bands corresponding to the $1\sigma$ confidence interval. The extracted valence-quark GPDs are found to be in good agreement with available lattice-QCD calculations. Our study demonstrates that DNN-based methods provide a flexible and robust framework for extracting pion GPDs and probing the multidimensional internal structure of the pion, offering a promising avenue for future investigations of hadron tomography.

\end{abstract}

\maketitle

\vspace{0.5em}

 \section{Introduction}
\label{intro}

One of the central goals of quantum chromodynamics (QCD) is to understand the three-dimensional structure of hadrons in terms of their quark and gluon degrees of freedom. Experimental observables such as parton distribution functions (PDFs), electromagnetic form factors (EMFFs), and distribution amplitudes provide complementary information on hadron structure. A unified description is provided by generalized parton distributions (GPDs) \cite{Muller:1994ses,Diehl:2003ny}, which simultaneously encode the longitudinal momentum and transverse spatial distributions of partons. In addition, GPDs provide access to the orbital angular momentum of partons through Ji's sum rule \cite{Ji:1996ek}, as well as to gravitational form factors (GFFs) and the mechanical properties of hadrons \cite{Burkert:2018bqq,Polyakov:2018zvc,Dwibedi:2026ozl}. Experimentally, GPDs can be accessed through hard exclusive processes such as deeply virtual Compton scattering and deeply virtual meson production. Extensive measurements of these processes have been performed by the H1 \cite{H1:2005gdw,H1:2007vrx,H1:2009wnw}, ZEUS \cite{ZEUS:2003pwh,ZEUS:2008hcd}, HERMES \cite{HERMES:2008abz,HERMES:2009cqe}, Jefferson Lab \cite{JeffersonLabHallA:2006prd,JeffersonLabHallA:2015dwe}, CLAS \cite{CLAS:2001wjj,CLAS:2008ahu}, and COMPASS \cite{COMPASS:2018pup} collaborations. Significant advances in mapping hadron structure are also anticipated from the future Electron-Ion Collider (EIC) at Brookhaven National Laboratory \cite{accardi2014electronioncolliderqcd}. Unlike PDFs, which describe only the longitudinal momentum structure of hadrons, GPDs depend on three kinematic variables: the longitudinal momentum fraction $x$, the skewness parameter $\xi$, and the squared momentum transfer $t$.

Among all hadrons, the pion occupies a unique position as the lightest hadron and the pseudo-Goldstone boson associated with the spontaneous breaking of chiral symmetry. Understanding its internal structure is therefore essential for elucidating the nonperturbative dynamics of QCD. Experimental information on the pion partonic structure has primarily been obtained from pion-induced Drell-Yan processes ($\pi^{\pm}N\rightarrow \ell^+\ell^-X$) \cite{Drell:1970wh}, pion-induced $J/\psi$ production ($\pi^{\pm}N\rightarrow J/\psi+X$) \cite{E672:1995won}, pion-induced prompt-photon production ($\pi^{\pm}N\rightarrow \gamma X$) \cite{WA70:1987bai}, leading-neutron production in deep-inelastic scattering ($ep\rightarrow e+n+X$) \cite{H1:2010hym}, pion electroproduction measurements ($ep\rightarrow e+n+\pi^{\pm}$) \cite{Brown:1973wr,Ackermann:1977rp,Brauel:1979zk,JeffersonLabFpi:2000nlc,JeffersonLabFpi-2:2006ysh,JeffersonLabFpi:2007vir,JeffersonLab:2008jve,Bebek:1977pe}, and pion elastic scattering ($\pi e\rightarrow \pi e$) \cite{Adylov:1977kj,Dally:1981ur,Dally:1982zk,NA7:1986vav}. These measurements have played a crucial role in constraining the pion PDFs and form factors (FFs). In contrast, direct experimental information on pion GPDs remains extremely limited.
Consequently, our understanding of the multidimensional structure of the pion encoded in its GPDs is considerably less developed than our knowledge of its PDFs and FFs. The forthcoming EIC is expected to open a new avenue for investigating pion GPDs through the Sullivan process \cite{Chavez:2023hbl,Chavez:2022tkf,Chavez:2021koz}.

However, the close connection between GPD, PDFs and FFs enables the construction of phenomenological GPD parameterizations that satisfy known theoretical constraints while reproducing the available experimental data \cite{Diehl:2004cx,Goharipour:2025zsw}. Such approaches provide valuable insight into the three-dimensional structure of the pion in the absence of direct GPD measurements. Compared with the nucleon, the pion possesses only a single leading-twist unpolarized GPD for both valence-quark and gluon cases, making it an especially attractive system for investigating hadron tomography \cite{Meissner:2008ay}. Moreover, the pion plays a fundamental role in nuclear binding and contributes significantly to the structure of nucleons through its meson cloud \cite{Thomas:2007bc}. Consequently, valence-quark pion GPDs have been investigated within numerous theoretical frameworks, including the Nambu-Jona-Lasinio model \cite{Zhang:2021uak}, chiral quark models \cite{Broniowski:2003rp}, double-distribution approaches \cite{Polyakov:1999gs}, light-front constituent quark models \cite{Frederico:2009fk,Puhan:2025kzz}, and lattice-QCD \cite{Ding:2024saz,Lin:2023gxz}. At the same time, theoretical and lattice-QCD studies of pion gluon GPDs remain rather limited. Phenomenological investigations of pion gluon GPDs have been carried out in Refs.~\cite{Chavez:2021llq,Kaur:2025gyr}. On the lattice side, however, only the gluon GFFs and the corresponding D-term have been determined \cite{Hackett:2023nkr,Shanahan:2018pib}. To the best of our knowledge, no direct lattice-QCD determination of the full $x$-dependent pion gluon GPDs is currently available.

Despite their importance, GPDs remain among the most challenging quantities to determine because they are not directly observable. Instead, they enter measurable quantities through Compton FFs, which involve convolution integrals over the partonic momentum fraction. Consequently, extracting GPDs from experimental observables constitutes a highly ill-posed inverse problem. In recent years, deep neural-networks (DNNs) have emerged as powerful tools for addressing such inverse problems and have been successfully applied to a broad range of hadron-structure studies \cite{Kumericki:2011rz,Watkins:2025apc,Panjsheeri:2026szi,Xu:2026lko,CaleroDiaz:2025luc,Le:2025swl}.

In this work, we present a DNN extraction of the pion unpolarized GPDs for both valence and gluon sector using the pion PDFs from the JAM21 \cite{Barry:2021osv} and xFitter \cite{Novikov:2020snp} analysis as input. The GPD is parameterized through a physics-informed neural-network (PINN) ansatz that combines the known PDF behavior with a trainable neural network (NN) component while enforcing the relevant theoretical constraints, including PDF sum rules and the normalization of the pion FFs. Our NN, consist of four inputs, with activation function SiLU \cite{elfwing2017sigmoid} with a single output of pion GPD. We employ the same NN architecture for the extraction of valence-quark and gluon GPDs through independent fits with different output layers tailored to each case. The network parameters are determined through the minimization of a $\chi^2$-based loss function that incorporates both experimental and lattice-QCD information. This framework enables a flexible and largely model-independent determination of pion GPDs together with reliable uncertainty estimates obtained from the full ensemble of PDF replicas.

The remainder of this paper is organized as follows. In Sec.~\ref{method}, we present the theoretical formalism, NN architecture, and training methodology. In Sec. ~\ref{nnfitting}, we discuss the NN fitting procedure in detail. Section \ref{sec:results} contains the numerical results and discussion, including the extracted pion GPDs and their comparison with experimental and lattice-QCD data. Finally, Sec.\ref{conclu} summarizes our conclusions and discusses future prospects.

\begin{figure}[t]
    \centering
    \includegraphics[width=\columnwidth]{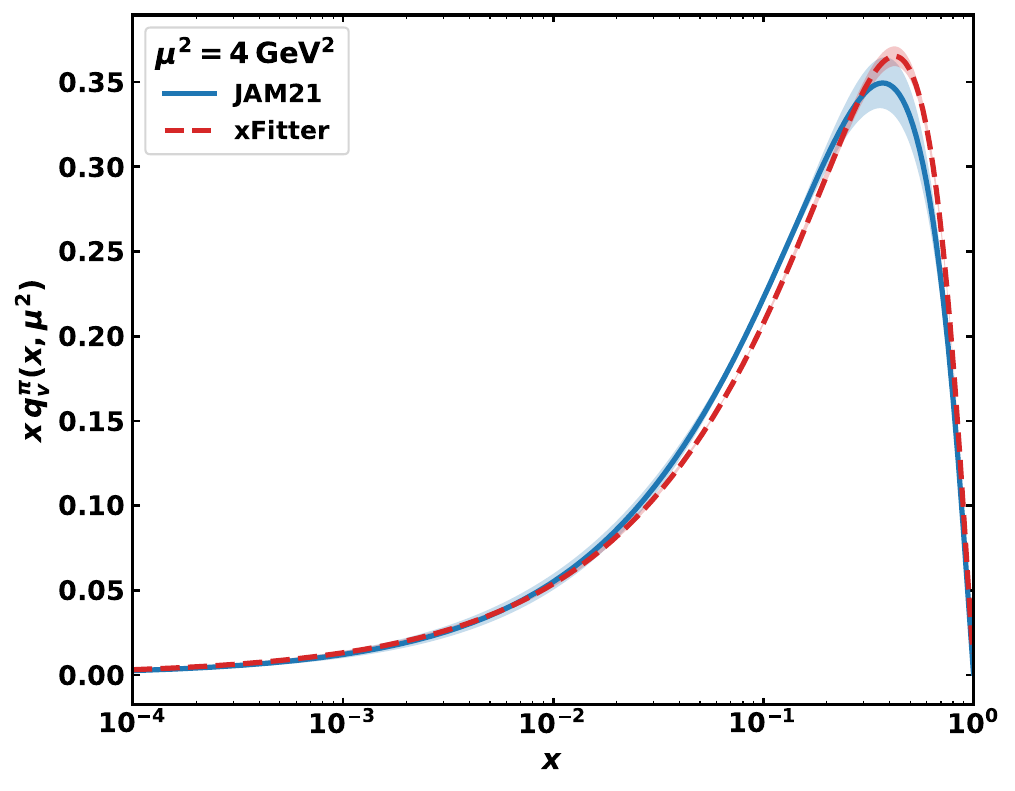}
    \caption{Valence-quark PDFs of the pion obtained using the JAM21 \cite{Barry:2021osv} and xFitter \cite{Novikov:2020snp} parameterizations at $\mu^2 = 4~\mathrm{GeV}^2$.}
    \label{fig:pdf_all_q2}
\end{figure}

\begin{figure}[t]
    \centering
    \includegraphics[width=\columnwidth]{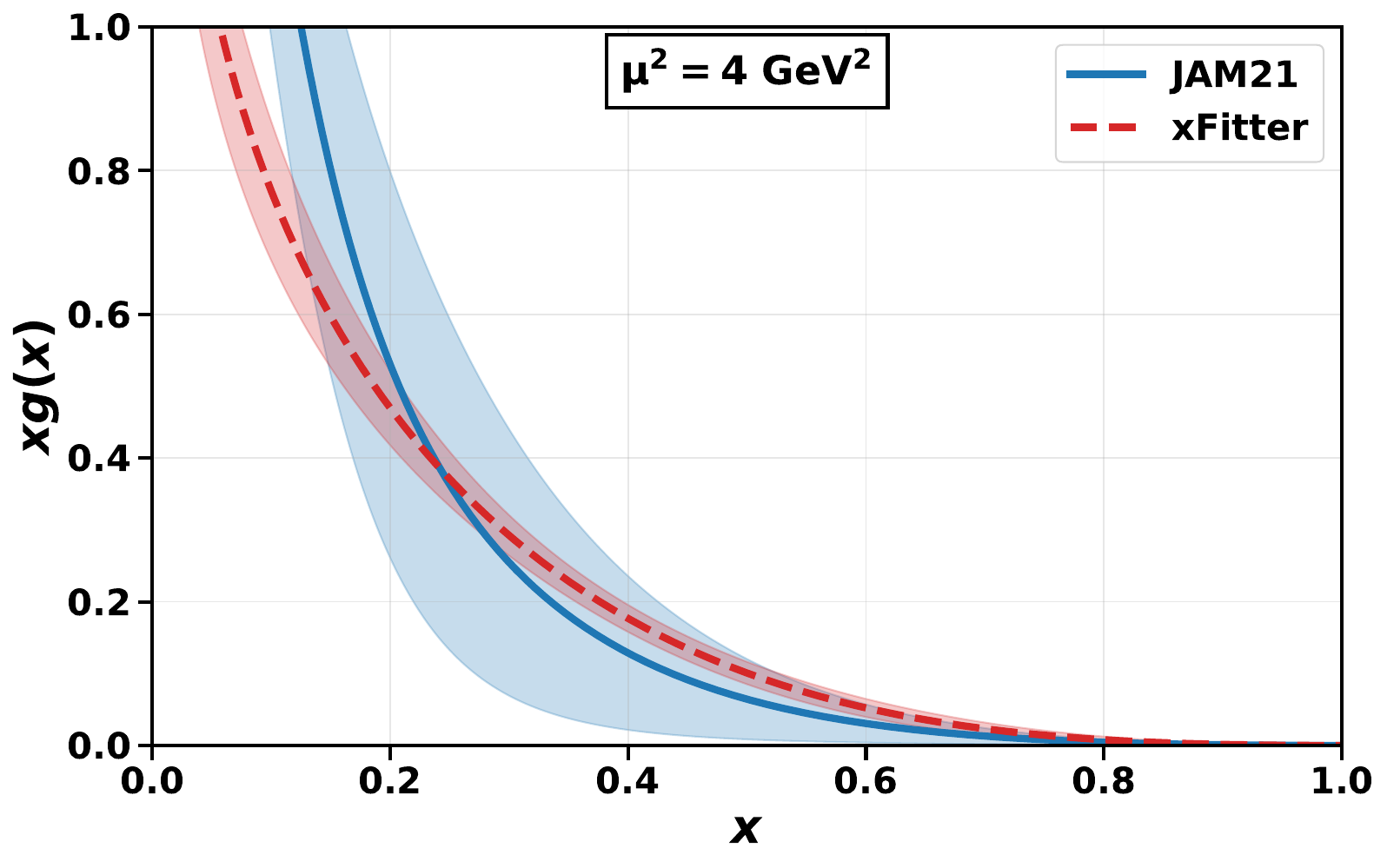}
    \caption{Gluon PDFs of the pion obtained using the central replica with a $1\sigma$ uncertainty band from the JAM21 \cite{Barry:2021osv} and xFitter \cite{Novikov:2020snp} parameterizations at $\mu^2 = 4~\mathrm{GeV}^2$.}
    \label{gluonplot}
\end{figure}

 \begin{figure*}[t]
    \centering
    \includegraphics[width=\textwidth]{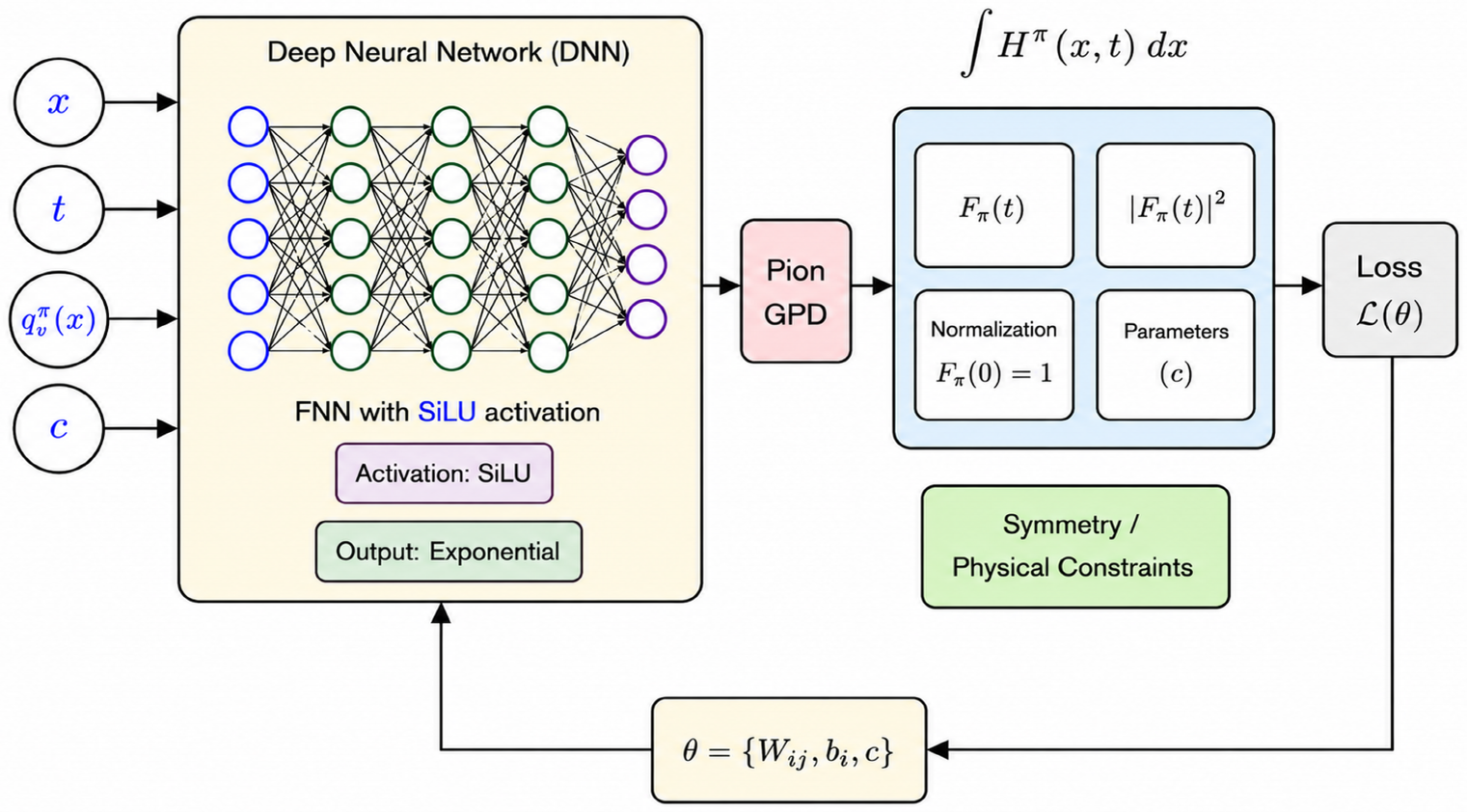}
    \caption{Schematic illustration of the NN framework used for the extraction of valence quark GPDs. Valence-quark PDFs, pion FFs data, and lattice-QCD inputs are used to train the network, while theoretical constraints such as charge conservation, normalization, and positivity are incorporated during the optimization procedure.}
    \label{DNN}
\end{figure*}
\begin{figure*}[t]
    \centering
    \includegraphics[width=0.49\textwidth]{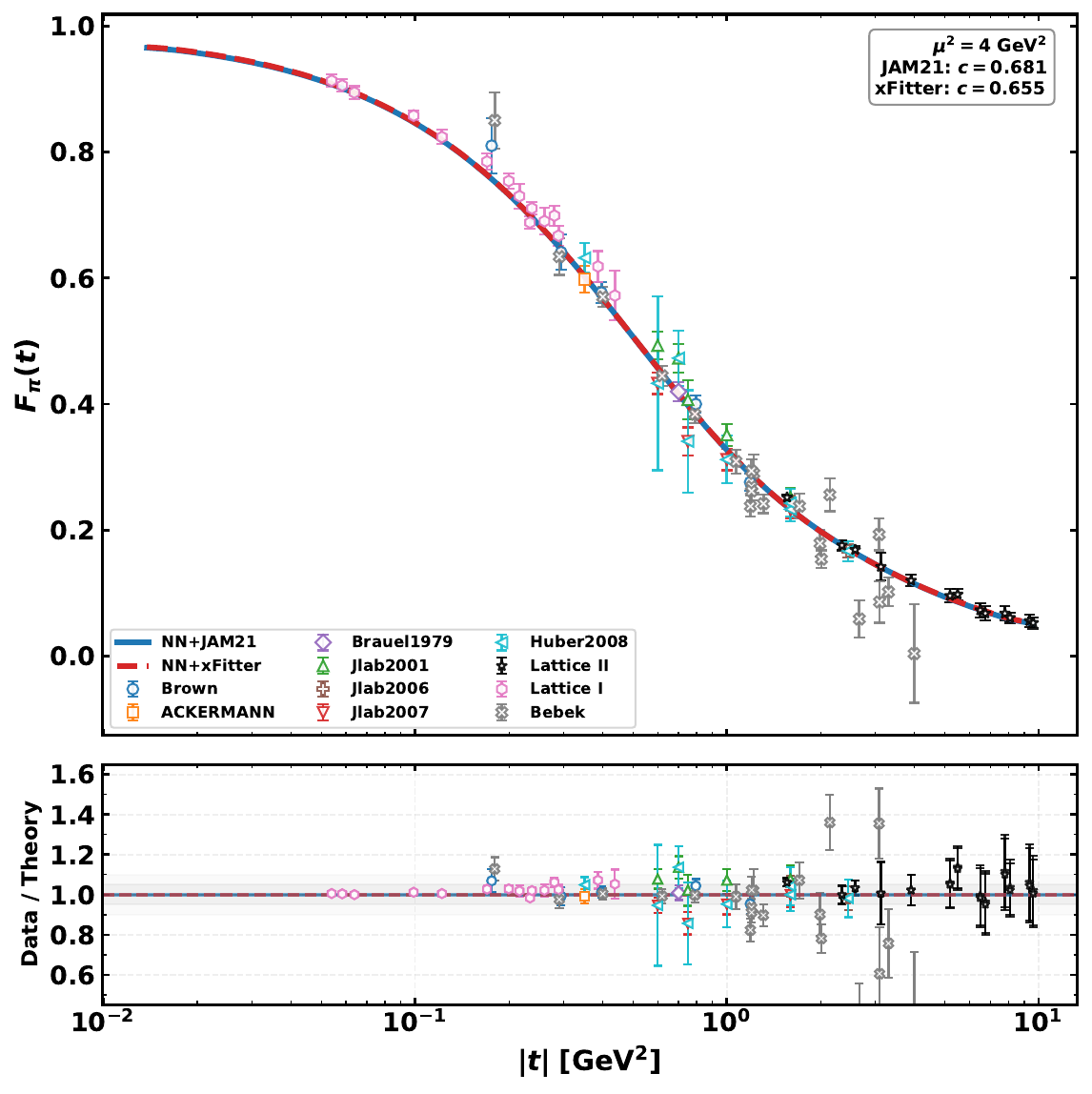}
    \hfill
    \includegraphics[width=0.49\textwidth]{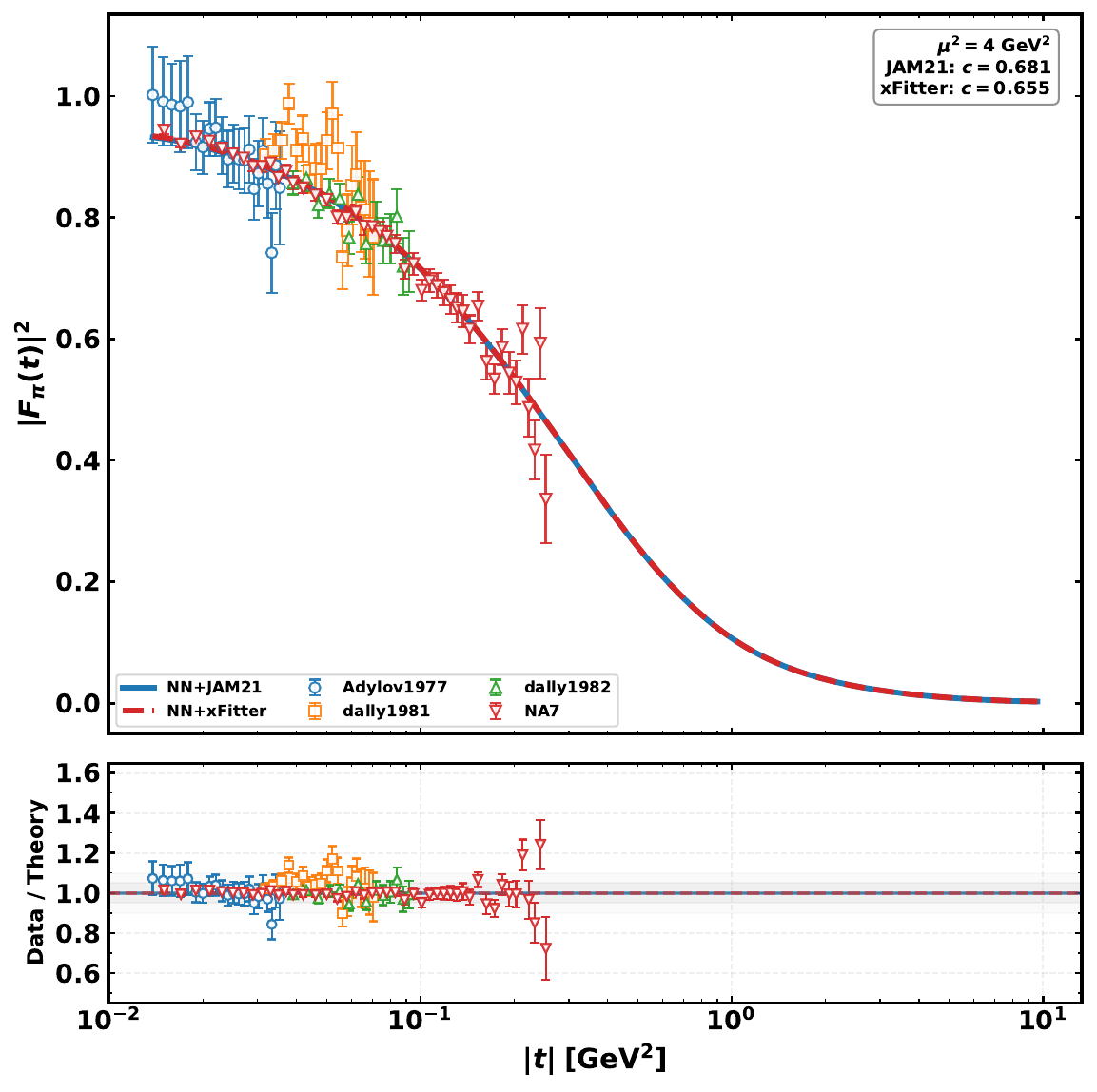}

    \caption{Comparison of the NN fits with the experimental
\cite{Brown:1973wr,Ackermann:1977rp,Brauel:1979zk,JeffersonLabFpi:2000nlc,JeffersonLabFpi-2:2006ysh,JeffersonLabFpi:2007vir,JeffersonLab:2008jve,Bebek:1977pe,Adylov:1977kj,Dally:1981ur,Dally:1982zk,NA7:1986vav}
and lattice-QCD data \cite{Gao:2021xsm,Ding:2024lfj} for the pion FF $F_\pi(t)$ (left panel) and its squared magnitude $|F_\pi(t)|^2$ (right panel), at $\mu^2 = 4~\mathrm{GeV}^2$. The fits are obtained using the JAM21 \cite{Barry:2021osv} and xFitter \cite{Novikov:2020snp} pion PDF parameterizations as inputs to the GPD framework.}
    \label{fig:fpi_q2_4}
\end{figure*}
\begin{figure}[t]
    \centering
    \includegraphics[width=\columnwidth]{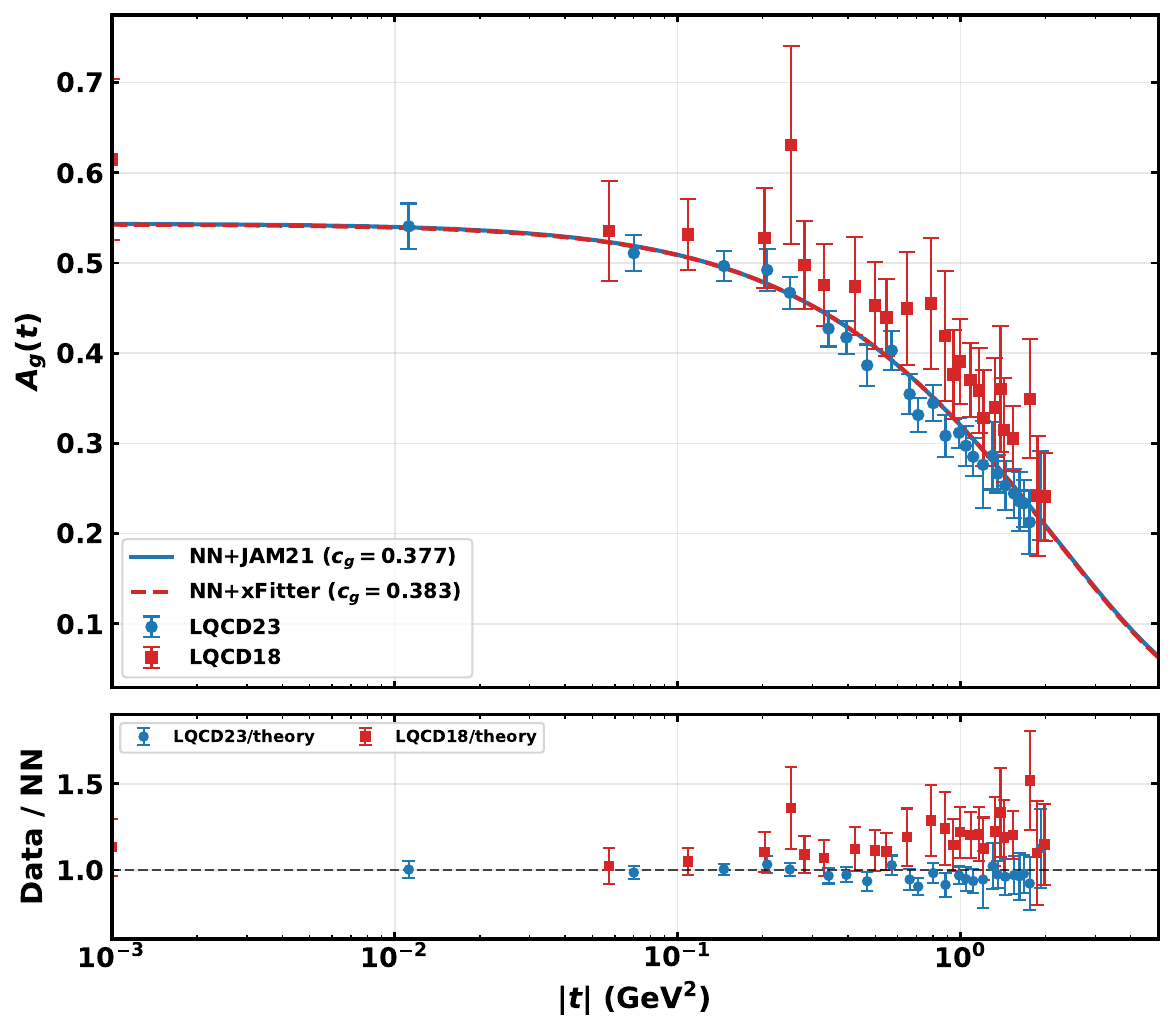}
    \caption{NN fit to the gluon GFF $A_g(t)$ using the available lattice-QCD results from Lattice23 \cite{Hackett:2023nkr} and Lattice18 \cite{Shanahan:2018pib}. The fit is performed using the JAM21 \cite{Barry:2021osv} and xFitter \cite{Novikov:2020snp} pion PDF parameterizations as inputs to the GPD framework.}
    \label{fig:Ag_fit}
\end{figure}
\begin{figure*}[t]
    \centering
    \includegraphics[width=\textwidth]{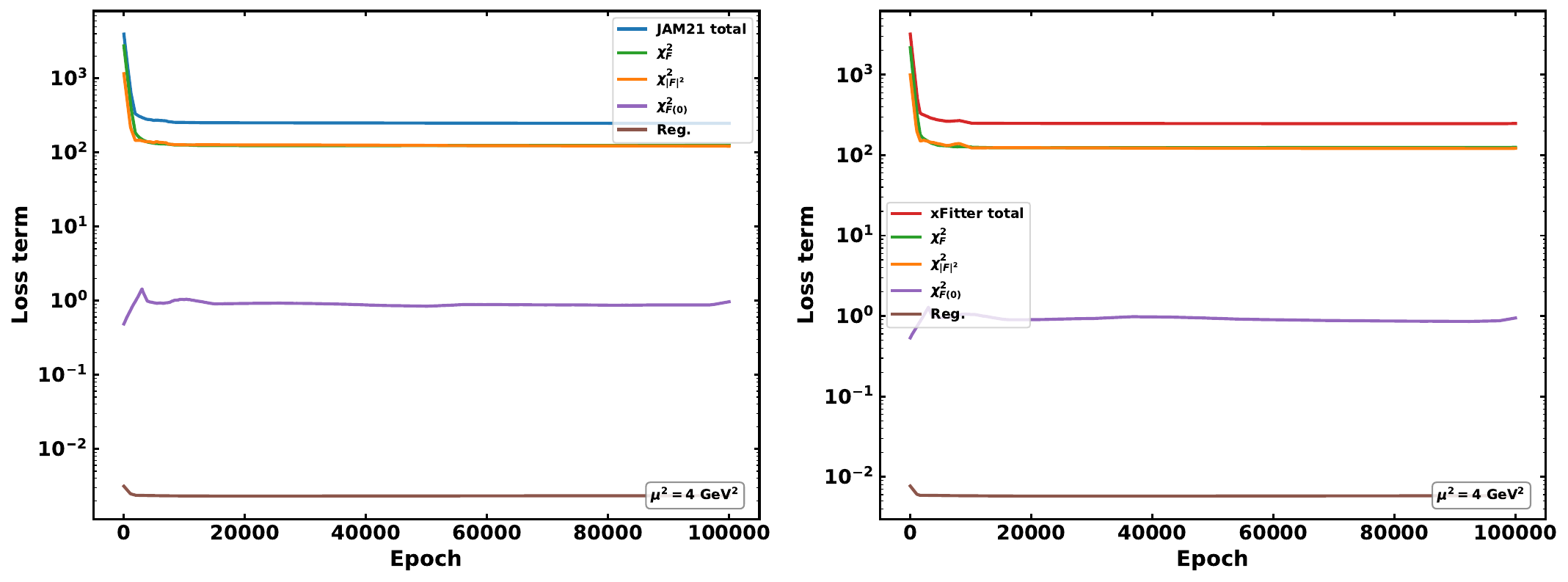}
 \caption{Evolution of the individual loss components and the total loss function as functions of the training epoch up to $10^{5}$ epochs at $\mu^2 = 4~\mathrm{GeV}^2$. The left and right panels correspond to the analysis using the JAM21 and xFitter pion PDFs, respectively. The observed convergence of all loss terms demonstrates the stability and robustness of the NN training procedure.}
    \label{fig:loss_components}
\end{figure*}
\begin{figure*}[t]
    \centering
    \includegraphics[width=\textwidth]{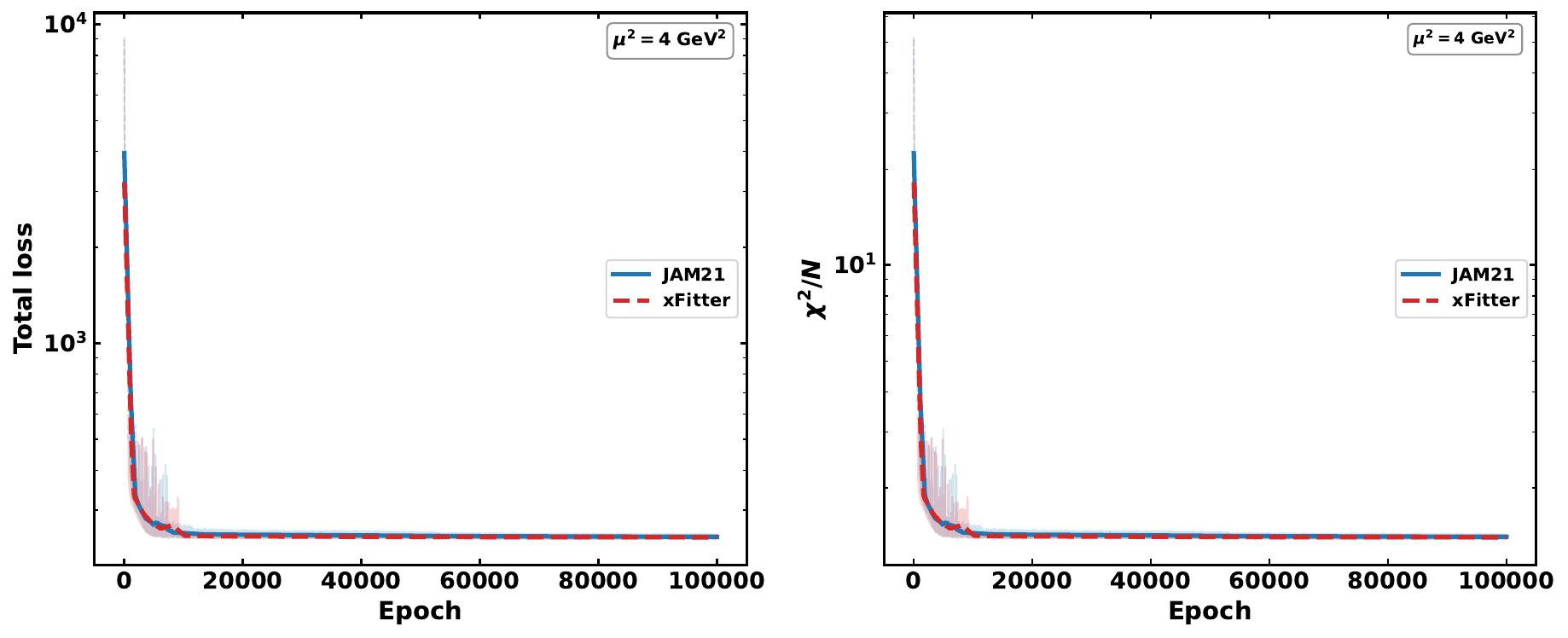}
    \caption{Training history of the total loss function and $\chi^2/N$ as functions of the training epoch up to $10^{5}$ epochs at $\mu^2 = 4~\mathrm{GeV}^2$ for the JAM21 and xFitter pion PDF analysis. The convergence behavior demonstrates the robustness of the NN optimization.}
    \label{fig:training_loss}
\end{figure*}
\begin{figure*}[t]
    \centering
    \includegraphics[width=0.49\textwidth]{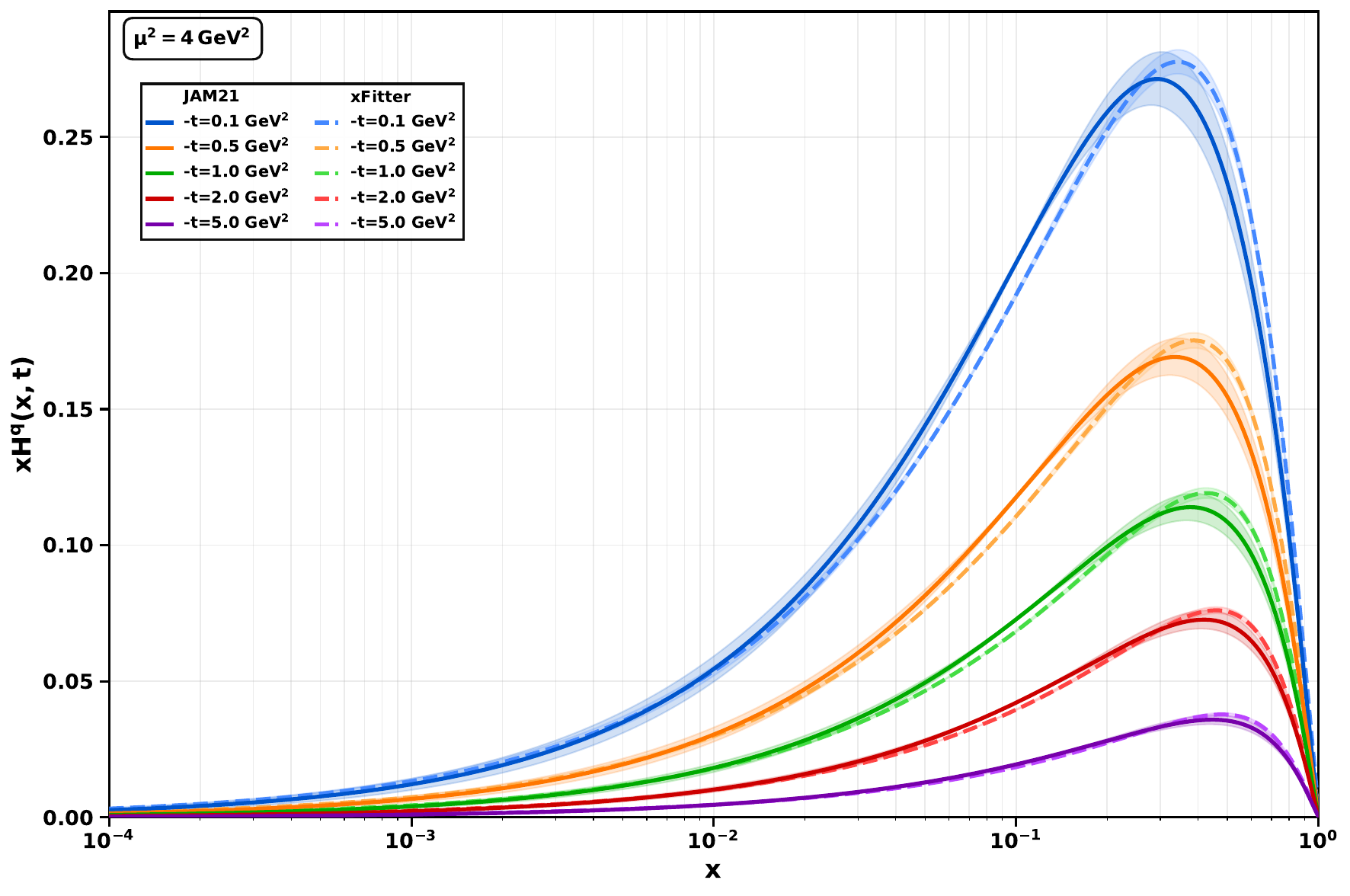}
    \hfill
    \includegraphics[width=0.49\textwidth]{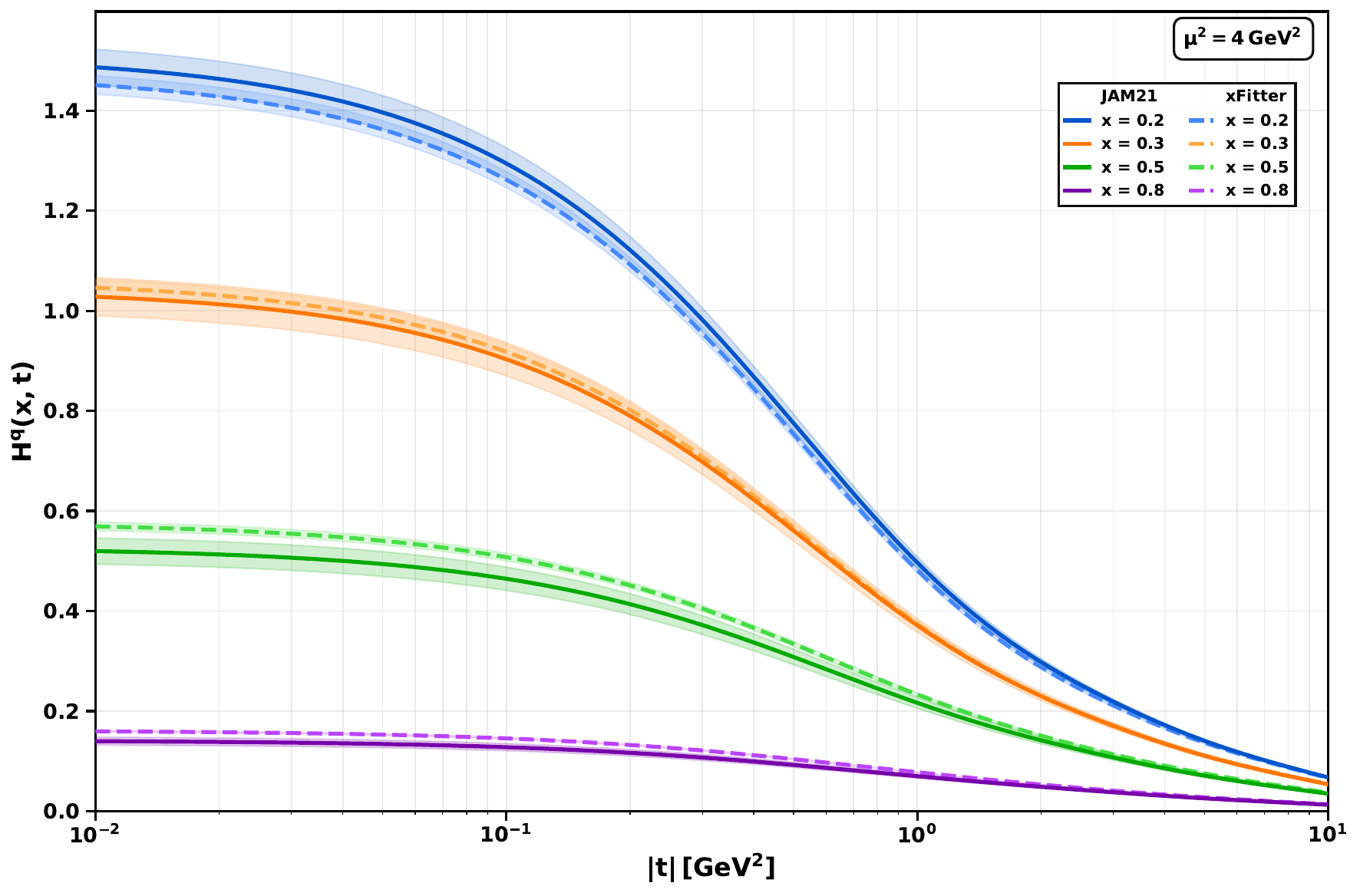}
    \caption{
    Comparison of the extracted quark GPDs obtained using the JAM21 and xFitter pion PDF inputs at the scale $\mu^2 = 4~\mathrm{GeV}^2$. The left panel shows $xH^{q}(x,t)$ as a function of the longitudinal momentum fraction $x$ for representative values of the momentum transfer $t$, while the right panel presents $H^{q}(x,t)$ as a function of $|t|$ for selected values of $x$. The shaded bands represent the uncertainties propagated from the valence-quark PDF fits.}
    \label{fig:gpd_q2_4}
\end{figure*}

\begin{figure*}[t]
    \centering
    \includegraphics[width=0.49\textwidth]{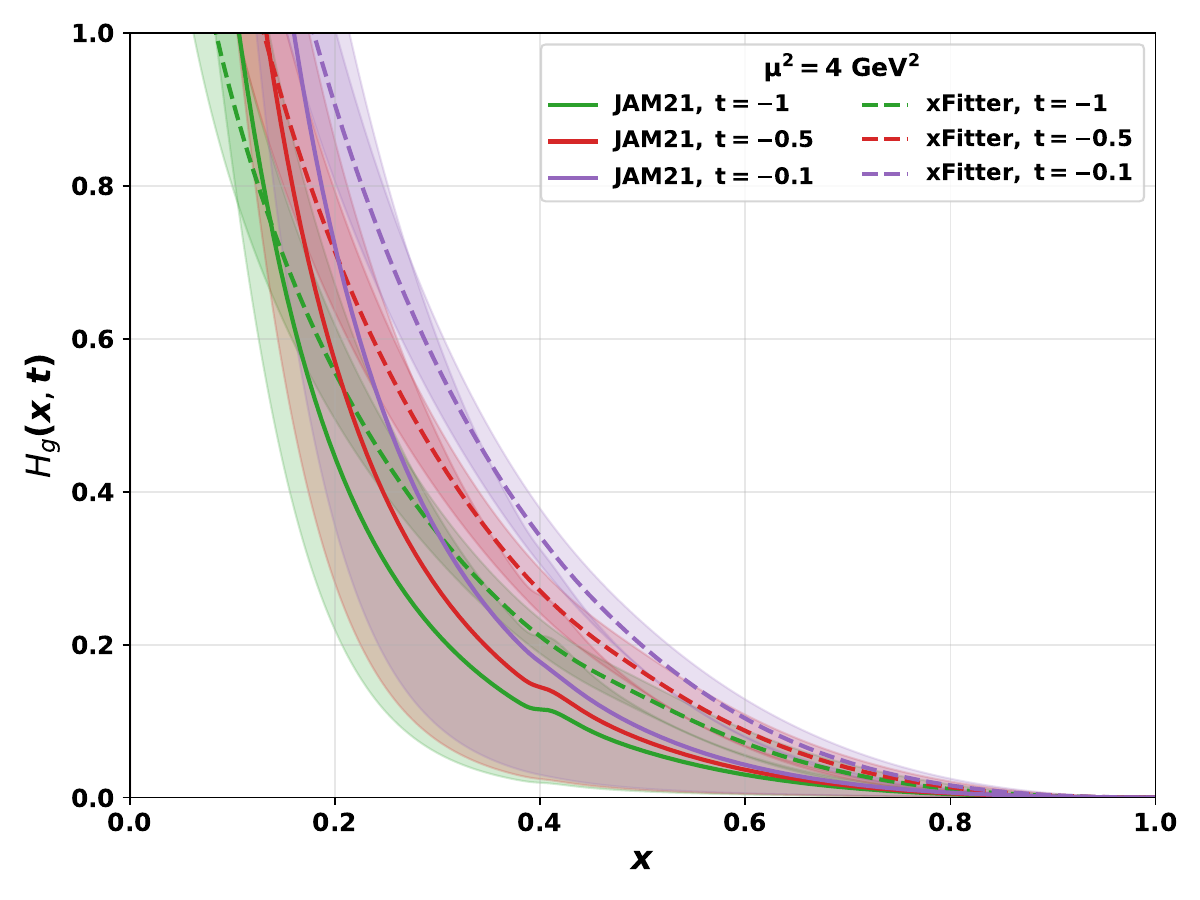}
    \hfill
    \includegraphics[width=0.49\textwidth]{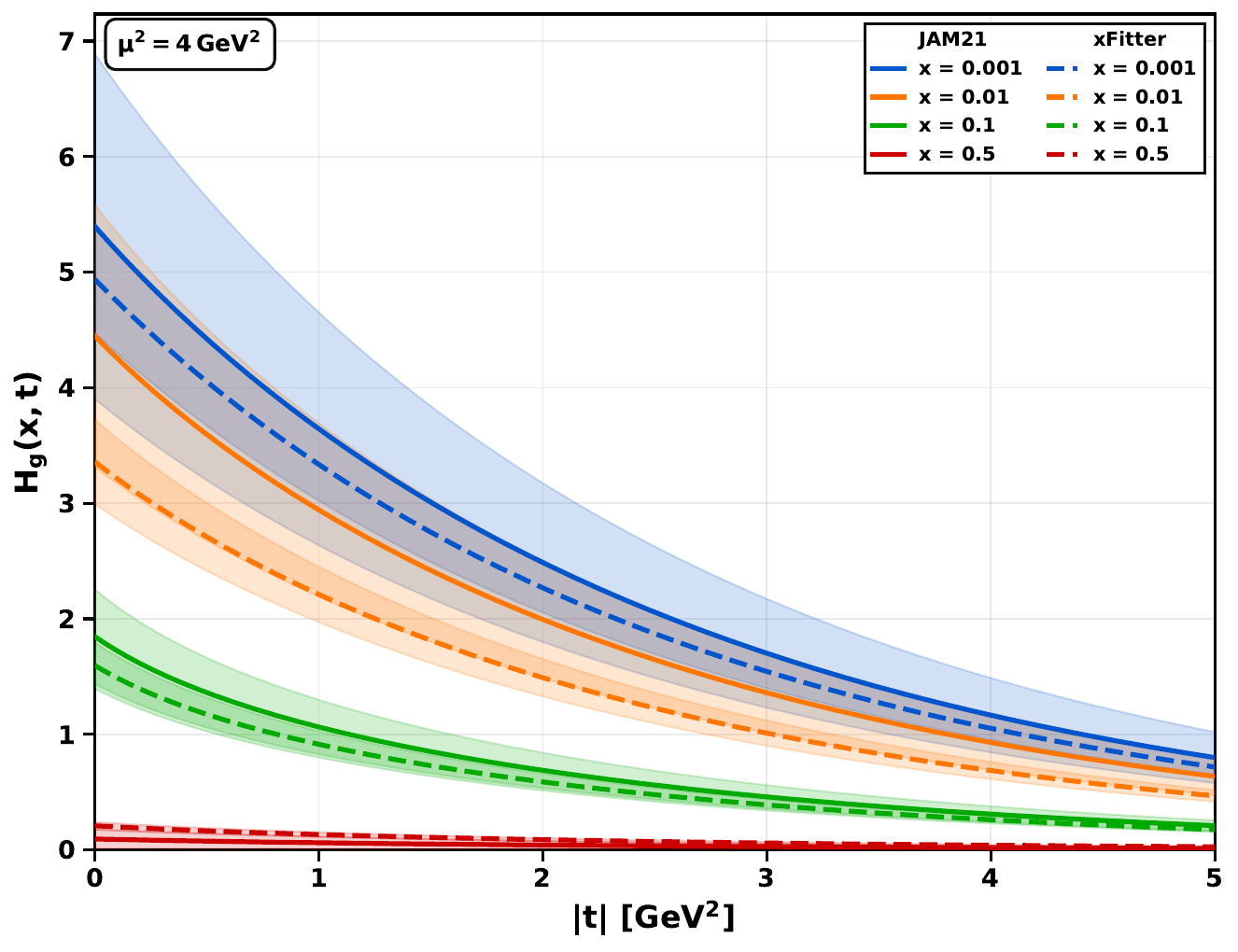}
    \caption{Gluon GPD extracted using the JAM21 and xFitter pion PDF parameterizations as inputs at the scale $\mu^2 = 4~\mathrm{GeV}^2$. The left panel shows the momentum-weighted gluon GPD, $H^{g}(x,t)$, as a function of the longitudinal momentum fraction $x$ for representative values of the momentum transfer $t$. The right panel presents the gluon GPD, $H^{g}(x,t)$, as a function of $|t|$ for selected values of $x$. The shaded bands represent the one-standard-deviation uncertainties propagated from the NN fit.}
    \label{fig:gluongpd_q2_4}
\end{figure*}
 \begin{figure*}[t]
    \centering
\includegraphics[width=\textwidth]{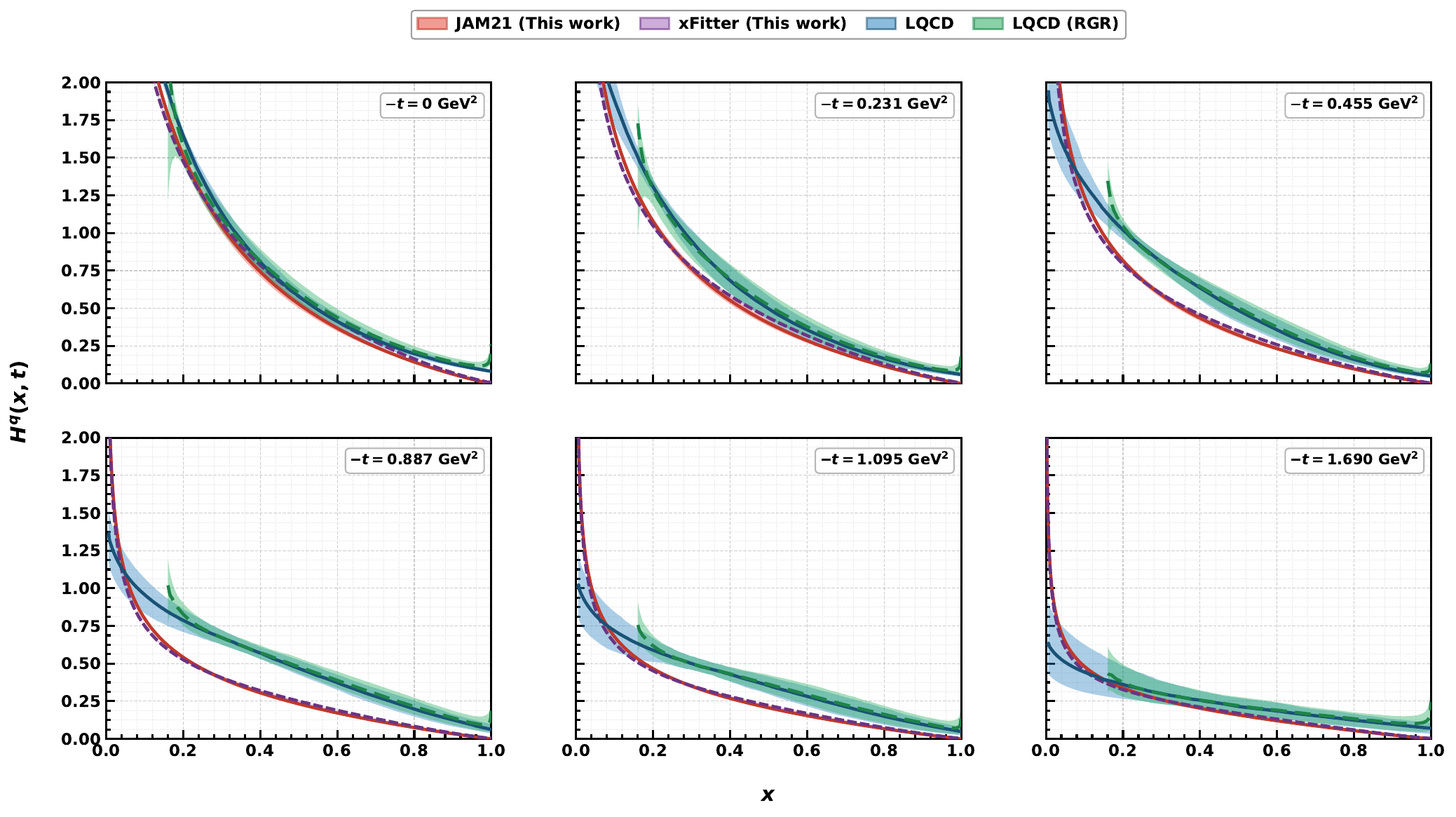}
    \caption{The calculated quark GPDs from the NN framework are shown as functions of the longitudinal momentum fraction $x$ at fixed values of $-t = 0$, $0.231$, $0.455$, $0.887$, $1.095$, and $1.690~\mathrm{GeV}^2$. The results are compared with the available lattice-QCD predictions of Ref.~\cite{Ding:2024saz}.}
    \label{fig8_lineargluon}
\end{figure*}
\section{Methodology\label{method}}
For spin-0 hadrons, there is only one leading-twist chiral-even unpolarized GPD for both quarks and gluons. This is in contrast to the nucleon, where two independent leading-twist unpolarized GPDs, $H$ and $E$, contribute. The pion quark and gluon GPDs are defined through the light-cone quark--quark and gluon--gluon correlation functions, respectively, as \cite{Diehl:2003ny}
 \begin{figure*}[t]
    \centering
\includegraphics[width=\textwidth]{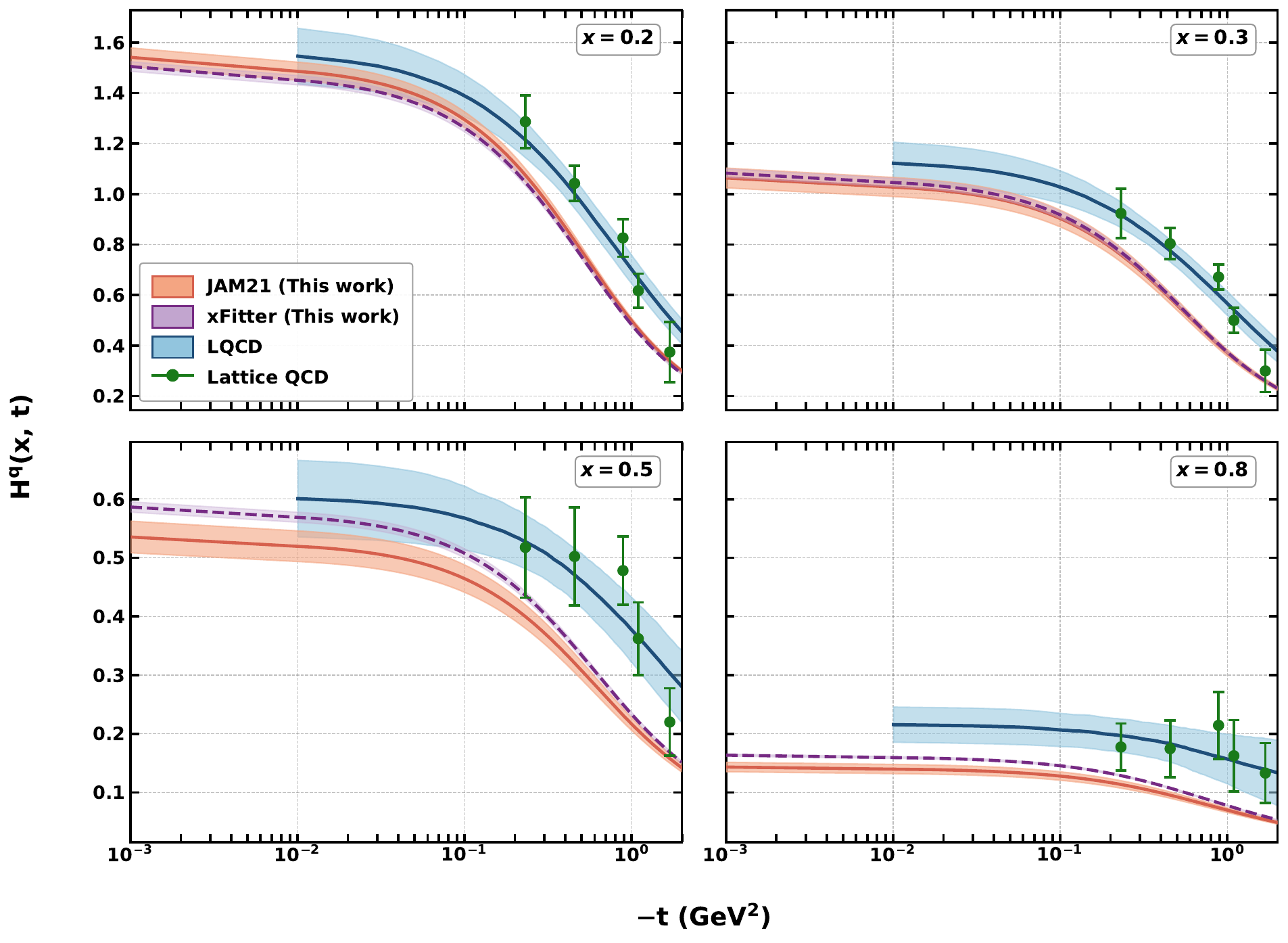}
 \caption{The calculated quark GPDs from the NN framework are shown as functions of the momentum transfer $-t$ (GeV$^2$) at fixed longitudinal momentum fractions $x = 0.2$, $0.3$, $0.5$, and $0.8$. The results are compared with the available lattice-QCD predictions of Ref.~\cite{Ding:2024saz}.}
\label{gpd_H_vs_t_log}
\end{figure*}
\begin{figure}[t]
    \centering
    \includegraphics[width=\columnwidth]{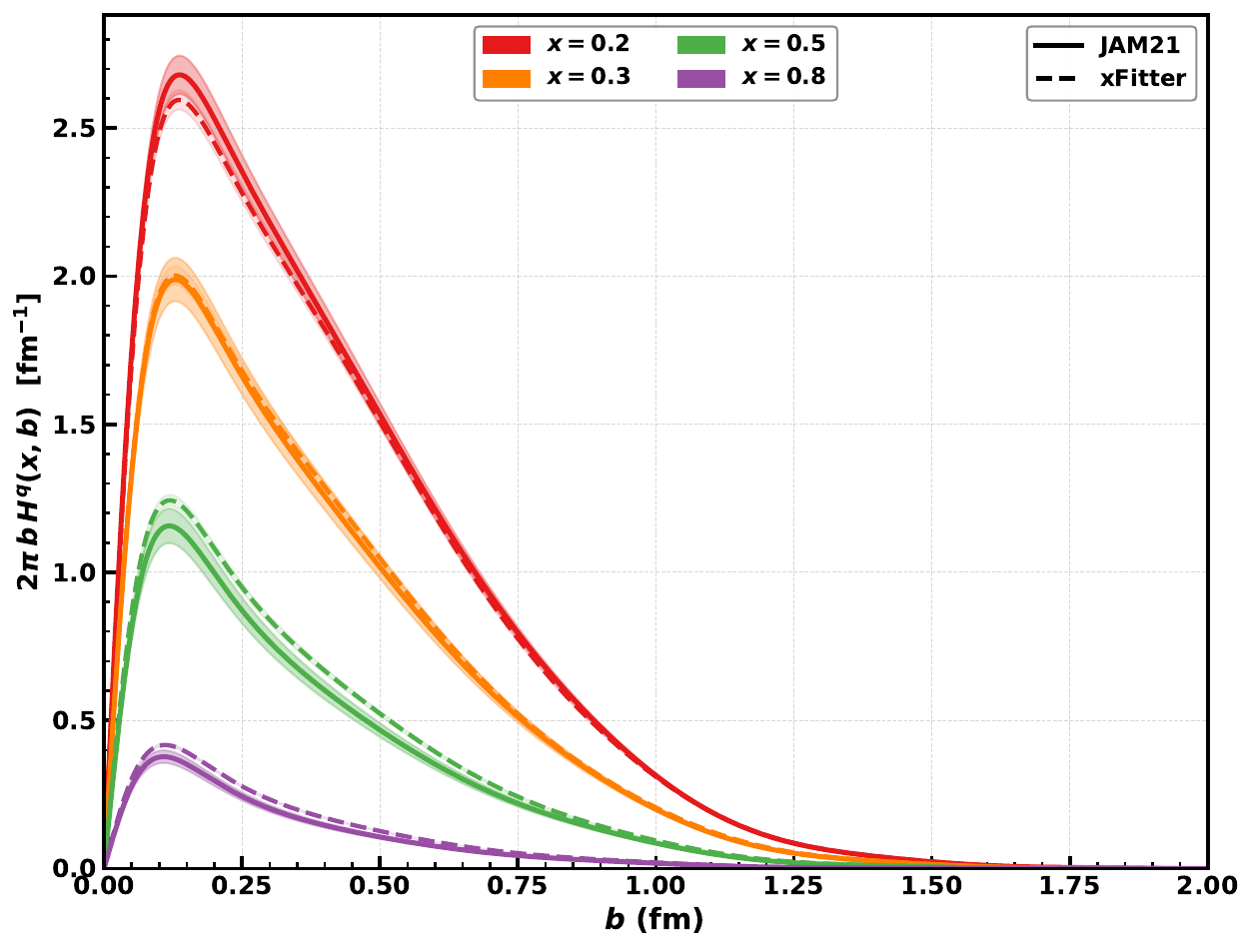}
    \caption{Impact-parameter dependent parton distribution function $ 2 \pi b H^q(x,\mathbf{b})$ with fixed values of $x$.
    %
    }
    \label{fig:Hq_density_vs_b}
\end{figure}
\begin{figure}[t]
    \centering
    \includegraphics[width=\columnwidth]{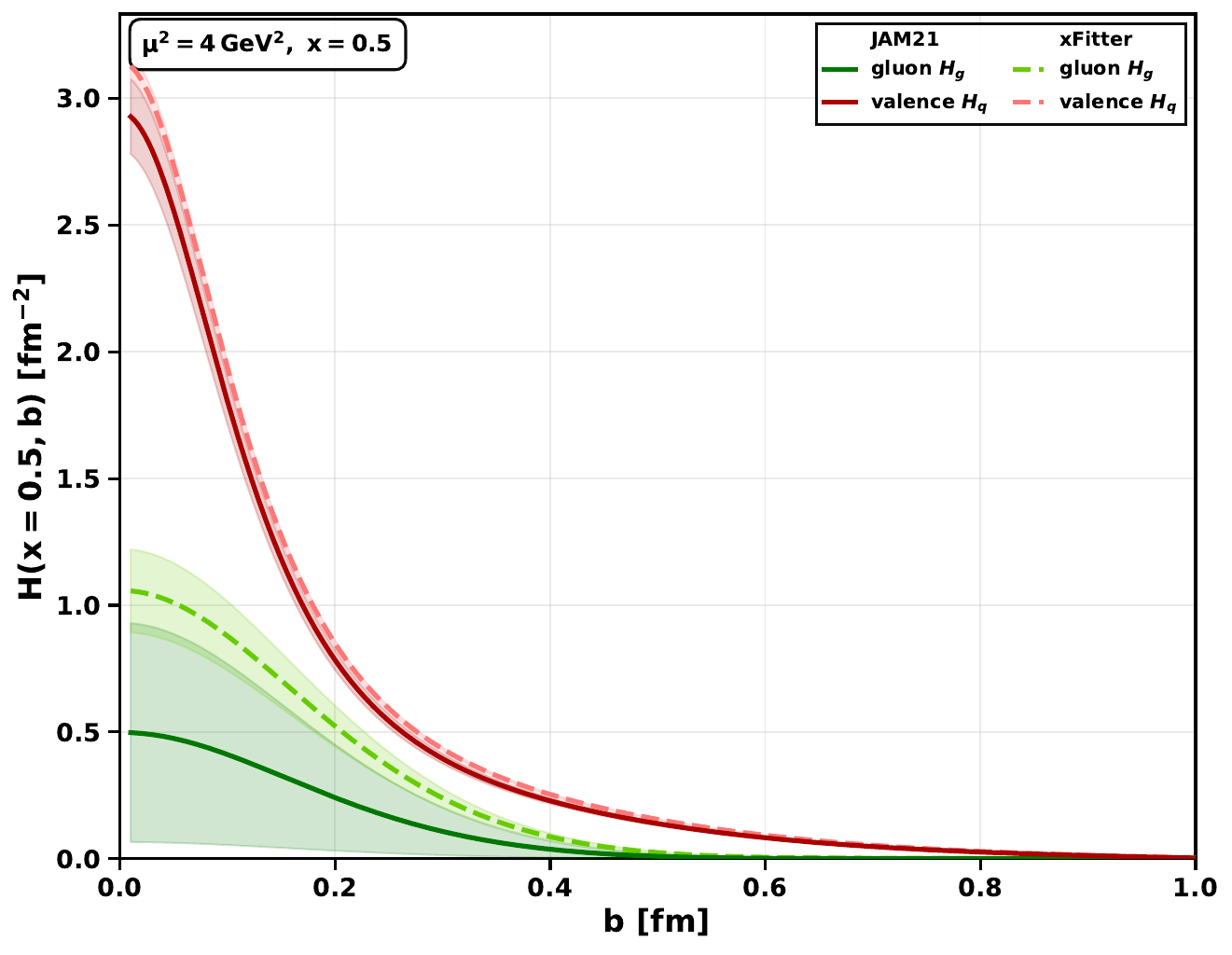}
    \caption{
    Impact-parameter dependent gluon distribution, $2\pi b\,H^{g}(x,\mathbf{b})$, as a function of the transverse impact parameter $b$ for fixed value of $x=0.5$ at the scale $\mu^2 = 4~\mathrm{GeV}^2$. The shaded bands denote the one-standard-deviation uncertainties propagated from the NN fit.
%
%
    }
    \label{fig:Hg_density_vs_b}
\end{figure}
\begin{figure*}[t]
    \centering
\includegraphics[width=\textwidth]{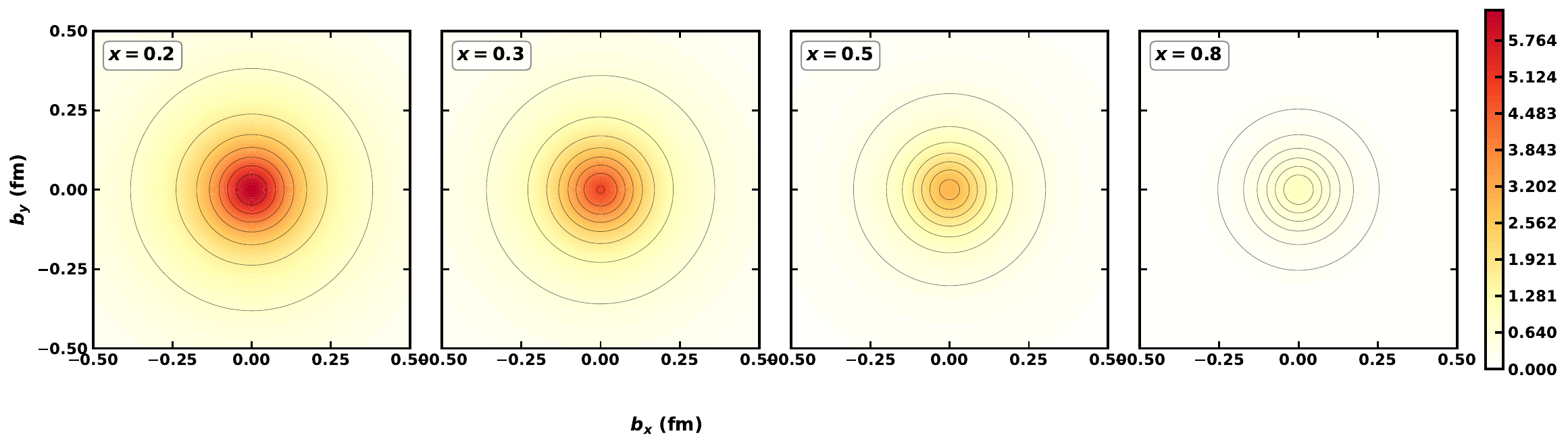}
    \caption{ Impact-parameter-dependent quark distribution, $2\pi b\,H^{q}(x,\mathbf{b})$, as a function of the transverse impact parameter $b$ for representative values of the longitudinal momentum fraction $x$, obtained using the JAM21 pion PDF parameterization at $\mu^2=4$ GeV$^2$. }
    \label{fig8_contour_JAM21}
\end{figure*}
\begin{figure*}[t]
    \centering
\includegraphics[width=\textwidth]{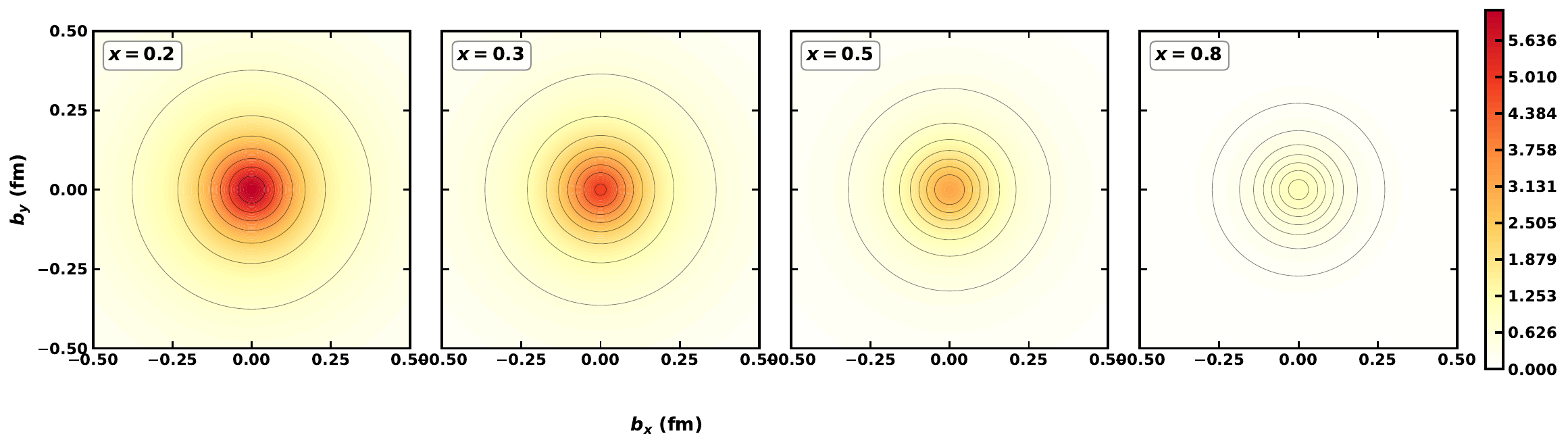}
    \caption{Impact-parameter-dependent quark distribution, $2\pi b\,H^{q}(x,\mathbf{b})$, as a function of the transverse impact parameter $b$ for representative values of the longitudinal momentum fraction $x$, obtained using the xFitter pion PDF parameterization at $\mu^2=4$ GeV$^2$. }
    \label{fig_contour_xfitter}
\end{figure*}

\begin{figure*}[t]
    \centering
    \includegraphics[width=\textwidth]{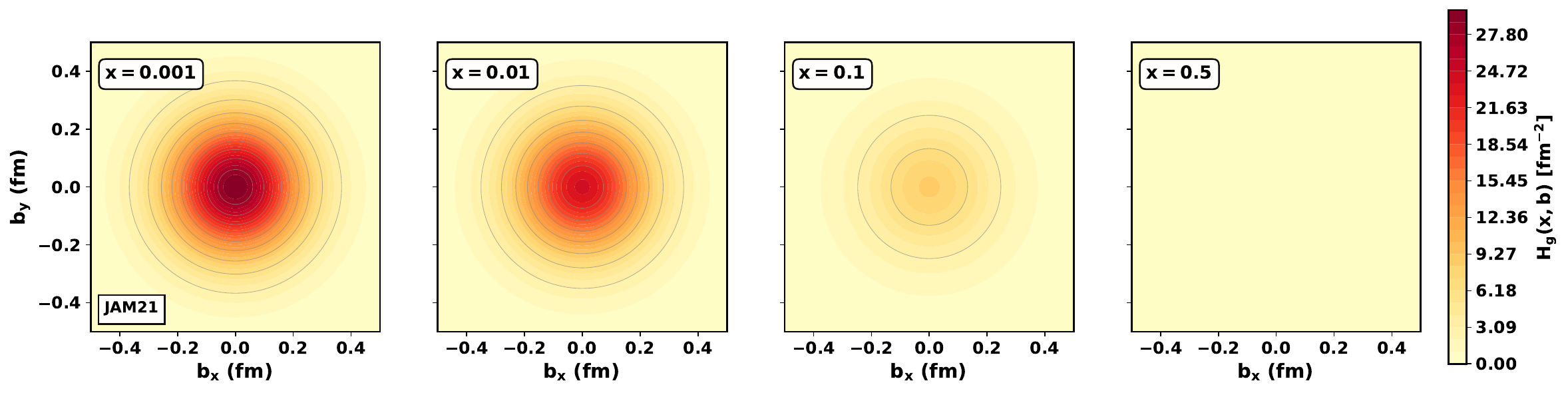}
    \caption{
    Two-dimensional transverse impact-parameter distributions of the gluon GPD, $H^{g}(x,\mathbf{b})$, in the $(b_x,b_y)$ plane for representative values of the longitudinal momentum fraction $x$, obtained using the JAM21 pion PDF parameterization at the scale $\mu^2 = 4~\mathrm{GeV}^2$.
    }
    \label{fig:contour_JAM21}
\end{figure*}
\begin{figure*}[t]
    \centering
    \includegraphics[width=\textwidth]{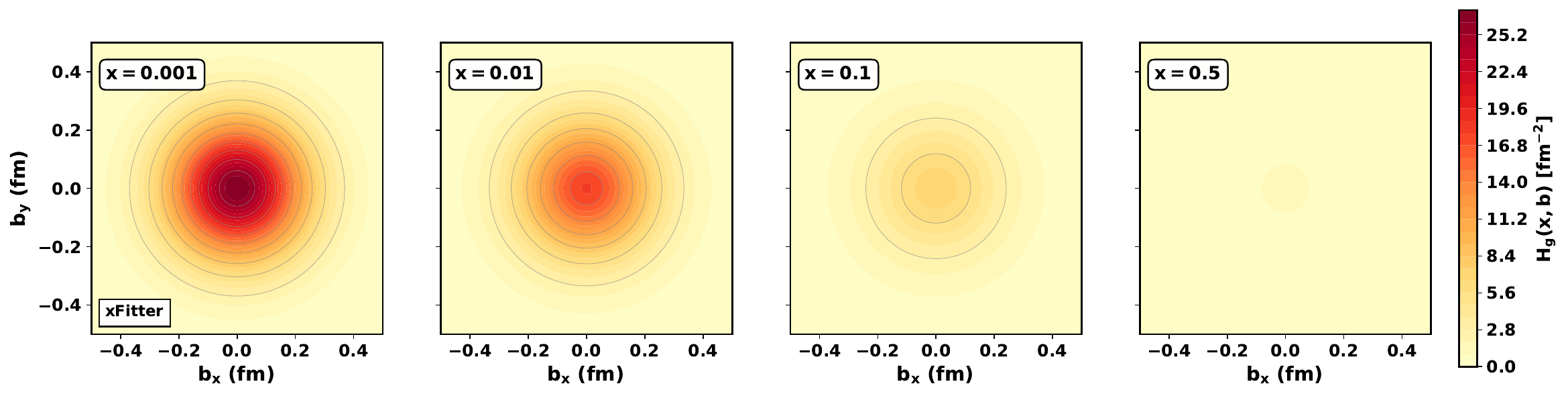}
    \caption{
    Two-dimensional transverse impact-parameter distributions of the gluon GPD, $H^{g}(x,\mathbf{b})$, in the $(b_x,b_y)$ plane for representative values of the longitudinal momentum fraction $x$, obtained using the xFitter pion PDF parameterization at the scale $\mu^2 = 4~\mathrm{GeV}^2$.
    }
    \label{fig:contour_xFitter}
\end{figure*}
\begin{equation}
\begin{aligned}
H_\pi^q(x,\xi,t)=&
\frac{1}{2}
\int\frac{dz^-}{2\pi}\,
e^{ixP^+z^-}
\\[2mm]
&\times
\left.
\langle\pi(P')|
\bar{\psi}\!\left(-\frac{z^-}{2}\right)
\gamma^+
\psi\!\left(\frac{z^-}{2}\right)
|\pi(P)\rangle
\right|_{z^+=\mathbf{z}_T=0},
\end{aligned}
\end{equation}
\begin{equation}
\begin{aligned}
H_\pi^g(x,\xi,t)=&
\frac{1}{P^+}
\int\frac{dz^-}{2\pi}\,
e^{ixP^+z^-}
\\[2mm]
&\times
\left.
\langle\pi(P')|
G^{+\mu}\!\left(-\frac{z^-}{2}\right)
G_{\mu}^{\ +}\!\left(\frac{z^-}{2}\right)
|\pi(P)\rangle
\right|_{z^+=\mathbf{z}_T=0},
\end{aligned}
\end{equation}
where $x$ denotes the longitudinal momentum fraction carried by the active quark or gluon, $\xi$ is the skewness parameter that characterizes the longitudinal momentum transfer, and $t=(P'-P)^2$ is the invariant momentum transfer squared.

Several phenomenological approaches have been proposed to model the pion GPDs in terms of collinear PDFs and EMFFs~\cite{Diehl:2004cx,Puhan:2026pkz,Goharipour:2025kif}. In the present work, we propose a PINN parameterization of the unpolarized pion quark and gluon GPDs in the zero-skewness ($\xi=0$) DGLAP region. The proposed framework reduces the model dependence associated with conventional multi-parameter functional forms while preserving the known theoretical constraints and the Regge-inspired behavior of the pion. The quark and gluon GPDs are parameterized as
\begin{equation}
\begin{aligned}
H^{q}(x,t) &= q_{v}^{\pi}(x)\,
e^{c|t|}\,
\mathrm{NN}(x,t),\\[2mm]
H_{g}(x,t) &= xg(x)\,
e^{c_{g}|t|}\,
\mathrm{NN}(x,t).
\end{aligned}
\end{equation}
where $q_v^\pi(x)$ and $xg(x)$ denote the pion valence-quark and gluon PDFs obtained from the JAM21 \cite{Barry:2021osv} and xFitter \cite{Novikov:2020snp} analysis, respectively. The parameters $c$ and $c_g$ are trainable Regge-inspired coefficients that govern the momentum-transfer dependence of the quark and gluon GPDs. The functions $\mathrm{NN}(x,t)$ represent the trainable NN components that encode the nonperturbative structure of the corresponding GPDs.

This parameterization satisfies the forward-limit constraint. In the limit $t\rightarrow0$, the quark and gluon GPDs reduce to their corresponding collinear pion PDFs
\begin{equation}
H^{q}(x,0)=q_{v}^{\pi}(x), \qquad
H^{g}(x,0)=xg(x),
\end{equation}
in accordance with the fundamental GPD sum rules.

For the present examination, we adopt the pion valence-quark and gluon PDFs from the JAM21 and xFitter global analysis. The JAM21 PDFs are primarily constrained by pion-induced Drell--Yan and leading-neutron production data, whereas the xFitter analysis additionally incorporates pion-induced prompt-photon production together with the Drell--Yan measurements. Consequently, both PDF sets provide phenomenologically well-constrained descriptions of the pion partonic structure and serve as reliable inputs for the NN framework developed in this work.

Throughout this examination, the pion valence-quark and gluon PDFs are evaluated at the scale
$\mu^2 = 4~\mathrm{GeV}^2$.
The corresponding valence-quark and gluon PDFs are shown in Figs.~\ref{fig:pdf_all_q2} and \ref{gluonplot}, respectively. The valence-quark PDFs obtained from the JAM21 and xFitter parameterizations exhibit very similar behavior over the entire $x$ range, providing stable and reliable inputs for the extraction of the quark GPD.

In contrast, the gluon PDFs exhibit noticeable differences, particularly in the small-$x$ region. The JAM21 gluon PDF is accompanied by a substantially larger uncertainty band, reflecting the broader spread among the LHAPDF replicas at low $x$, whereas the xFitter gluon PDF exhibits comparatively smaller uncertainties over the same kinematic region \cite{Buckley:2014ana}. Throughout this work, we employ the central replica of each PDF set together with its corresponding $1\sigma$ uncertainty band. These uncertainties are propagated through the fitting procedure to estimate the uncertainties of the extracted quark and gluon GPDs.

For arbitrary value of the skewness parameter, the first Mellin moment of the quark GPD is related to the pion EMFF through the sum rule as \cite{Goharipour:2025zsw}

\begin{equation}
F_\pi(t)
=
\sum_q e_q
\int_{-1}^{1} dx \,
H^q(x,\xi=0,t),
\label{eq:ff_sumrule}
\end{equation}
where $e_q$ denotes the electric charge of quark flavor $q$. At zero skewness, the second Mellin moment of the gluon GPD is related to the gluon GFF through the sum rule
\begin{equation}
A_{g}(t)
=
\int_{0}^{1} dx \,
x H_{g}(x,\xi=0,t),
\label{eq:gluon_sumrule}
\end{equation}
where $A_{g}(t)$ denotes the gluon contribution to the pion GFF. More generally, at nonzero skewness, the first Mellin moment satisfies the polynomiality relation
\begin{equation}
\int_{0}^{1} dx \,
H_{g}(x,\xi,t)
=
A_{g}(t)
+\xi^{2}D_\pi^{g}(t),
\end{equation}
where $D_\pi^{g}(t)$ denotes the gluon $D$-term GFF.

The fit is performed over the momentum-transfer range
\[
0.0138 \leq |t| \leq 9.77~\mathrm{GeV}^2,
\]
thereby probing both the long- and short-distance structure of the pion. To constrain the extraction of the pion valence-quark GPD, we employ a combined dataset consisting of experimental measurements and lattice-QCD calculations of the pion EMFF.

The experimental dataset includes pion electroproduction measurements
\[
ep \rightarrow e^{\prime}\pi^{+}n,
\]
from Refs.~\cite{Brown:1973wr,Ackermann:1977rp,Brauel:1979zk,JeffersonLabFpi:2000nlc,JeffersonLabFpi-2:2006ysh,JeffersonLabFpi:2007vir,JeffersonLab:2008jve,Bebek:1977pe}, together with elastic pion scattering data
\[
e^{-}\pi \rightarrow e^{-}\pi,
\]
from Refs.~\cite{Adylov:1977kj,Dally:1981ur,Dally:1982zk,NA7:1986vav}. In addition, two recent lattice-QCD determinations of the pion EMFF from Refs.~\cite{Gao:2021xsm,Ding:2024lfj} are incorporated into the analysis.

The complete dataset comprises 47 pion electroproduction, $101$ elastic pion scattering, and $28$ lattice-QCD data points, resulting in a total of $176$ data points. For the extraction of the gluon GPD, we employ the available lattice-QCD calculations of the pion gluon GFF, $A_g(t)$, from Refs.~\cite{Shanahan:2018pib,Hackett:2023nkr}. These datasets comprise 50 lattice-QCD data points spanning a broad range of momentum transfer. Such kinematic coverage provides stringent constraints on the momentum-transfer dependence of the gluon GFF and enables a robust determination of the pion gluon GPD within the proposed NN framework.

The parameters of the DNN are determined by minimizing a physics-informed loss function that incorporates the experimental and lattice-QCD data together with normalization and regularization constraints. 

\begin{table*}[t]
\caption{
Individual and total $\chi^2/N$ values obtained from the NN fit using the JAM21 and xFitter pion PDF parameterizations at $\mu^2 = 4~\mathrm{GeV}^2$. The individual contributions correspond to the experimental data from Refs.~\cite{Brown:1973wr,Ackermann:1977rp,Brauel:1979zk,JeffersonLabFpi:2000nlc,JeffersonLabFpi-2:2006ysh,JeffersonLabFpi:2007vir,JeffersonLab:2008jve,Bebek:1977pe,Adylov:1977kj,Dally:1981ur,Dally:1982zk,NA7:1986vav} and the lattice-QCD calculations from Refs.~\cite{Gao:2021xsm,Ding:2024lfj}.
}
\label{tab:chi2_dataset}
\centering
\begin{ruledtabular}
\begin{tabular}{lcccccc}
\multirow{2}{*}{Data set} &
\multirow{2}{*}{$N$} &
\multicolumn{2}{c}{JAM21} &
\multicolumn{2}{c}{xFitter} &
\multirow{2}{*}{Ref.} \\
\cline{3-6}
& & $\chi^2$ & $\chi^2/N$ & $\chi^2$ & $\chi^2/N$ & \\
\hline

Brown        &  5  &  3.97 & 0.79 &  3.92 & 0.78 &
\cite{Brown:1973wr} \\

Ackermann    &  1  &  0.09 & 0.09 &  0.10 & 0.10 &
\cite{Ackermann:1977rp} \\

Brauel       &  1  &  0.04 & 0.04 &  0.03 & 0.03 &
\cite{Brauel:1979zk} \\

JLab (2001)  &  5  & 11.33 & 2.27 & 11.16 & 2.23 &
\cite{JeffersonLabFpi:2000nlc} \\

JLab (2006)  &  2  &  0.81 & 0.41 &  0.79 & 0.39 &
\cite{JeffersonLabFpi-2:2006ysh} \\

JLab (2007)  &  4  & 10.08 & 2.52 & 10.25 & 2.56 &
\cite{JeffersonLabFpi:2007vir} \\

Huber (2008) &  8  &  4.02 & 0.50 &  3.98 & 0.50 &
\cite{JeffersonLab:2008jve} \\

Lattice II   & 13  & 12.50 & 0.96 & 12.01 & 0.92 &
\cite{Ding:2024lfj} \\

Lattice I    & 15  & 21.03 & 1.40 & 20.32 & 1.35 &
\cite{Gao:2021xsm} \\

Bebek        & 21  & 63.71 & 3.03 & 63.87 & 3.04 &
\cite{Bebek:1977pe} \\

Adylov       & 22  & 10.48 & 0.48 & 10.32 & 0.47 &
\cite{Adylov:1977kj} \\

Dally (1981) & 20  & 45.04 & 2.25 & 43.34 & 2.17 &
\cite{Dally:1981ur} \\

Dally (1982) & 14  &  8.83 & 0.63 &  9.03 & 0.65 &
\cite{Dally:1982zk} \\

NA7          & 45  & 57.28 & 1.27 & 57.06 & 1.27 &
\cite{NA7:1986vav} \\

\hline
\textbf{Total}
& \textbf{176}
& \textbf{249.22}
& \textbf{1.42}
& \textbf{246.17}
& \textbf{1.40}
& -- \\

\end{tabular}
\end{ruledtabular}
\end{table*}
\section{neural-network Fitting} 
\label{nnfitting}
In the present work, we employ a feed-forward DNN \cite{goodfellow2016deep,haykin2009neural} to extract the pion GPDs by simultaneously fitting the experimental pion EMFF and the lattice-QCD gluon GFF. The same NN architecture is adopted for both the valence-quark and gluon GPDs; however, the two networks are trained independently using different input datasets and physical constraints. A schematic representation of the valence-quark framework is shown in Fig.~\ref{DNN}.

The network consists of three hidden layers, each containing $64$ neurons. For the valence-quark GPD, the input variables are the longitudinal momentum fraction $x$ and the momentum transfer squared $t$. For the gluon GPD, the network additionally includes $\log x$ as an input feature to better capture the rapidly varying behavior of the gluon distribution in the small-$x$ region. In both cases, the corresponding pion PDFs are taken from the JAM21 and xFitter parameterizations. The Regge-inspired parameters controlling the momentum-transfer dependence of the GPDs, denoted by $c$ for the valence-quark case and $c_g$ for the gluon case, are treated as trainable parameters and are optimized simultaneously with the NN weights and biases during the training procedure.

The hidden layers employ the Sigmoid Linear Unit (SiLU) activation function \cite{elfwing2017sigmoid}
\begin{equation}
{\rm SiLU}(z)=z\,\sigma(z),
\end{equation}
where
\begin{equation}
\sigma(z)=\frac{1}{1+e^{-z}}
\end{equation}
where $\sigma(z)$ is the sigmoid function. The SiLU activation provides a smooth and continuously differentiable nonlinear mapping, leading to improved numerical stability during training while avoiding sharp discontinuities in the extracted distributions. Such smooth activation functions are particularly well suited for modeling continuous physical observables, including GPDs and FFs.

For the output layer, an exponential activation function is employed to ensure that the NN correction remains positive. Consequently, the extracted quark and gluon GPDs
\begin{equation}
H^{q}(x,t)\geq 0, \qquad
H_{g}(x,t)\geq 0,
\end{equation}
remain non-negative provided that the corresponding input PDFs are positive. The exponential output activation therefore incorporates a physically motivated positivity constraint directly into the NN framework. The exact form of the output layer have been discussed in Appendix \ref{aper}.

The trainable parameters of the valence-quark network are determined by minimizing the total loss function
\begin{equation}
\mathcal{L}_{\rm total}
=
\chi^2_{F_\pi}
+
\chi^2_{|F_\pi|^2}
+
\chi^2_{\rm norm}
+
\mathcal{L}_{\rm reg},
\end{equation}
where the individual contributions encode both the experimental information and the theoretical constraints.

The contribution associated with the experimental pion EMFF data is given by
\begin{equation}
\chi^2_{F_\pi}
=
\sum_i
\left[
\frac{
F_\pi^{\rm model}(t_i)
-
F_\pi^{\rm exp}(t_i)
}{
\sigma_i
}
\right]^2,
\end{equation}
where $F_\pi^{\rm model}(t_i)$ and $F_\pi^{\rm exp}(t_i)$ denote the theoretical prediction and the corresponding experimental measurement of the pion EMFF at momentum transfer $t_i$, respectively, while $\sigma_i$ represents the experimental uncertainty.

To further constrain the extraction, measurements of the squared pion EMFF are incorporated through
\begin{equation}
\chi^2_{|F_\pi|^2}
=
\sum_i
\left[
\frac{
|F_\pi^{\rm model}(t_i)|^2
-
|F_\pi^{\rm exp}(t_i)|^2
}{
\sigma_i
}
\right]^2.
\end{equation}

The FF sum rule condition

\begin{equation}
F_\pi(0)=1,
\end{equation}

is imposed through the normalization penalty
\begin{equation}
\chi^2_{\rm norm}
=
\left[
\frac{
F_\pi(0)-1
}{
0.01
}
\right]^2.
\end{equation}
To improve the numerical stability of the optimization and suppress unphysical oscillations in the extracted GPDs, an additional regularization term is introduced
\begin{equation}
\mathcal{L}_{\rm reg}
=
\lambda
\left\langle
F_\pi^2
\right\rangle,
\end{equation}
where $\lambda$ denotes the regularization strength. This term penalizes excessively large fluctuations in the NN output, thereby improving the stability of the optimization and ensuring a smooth and physically consistent behavior of the extracted pion GPD and the corresponding EMFF.

For the gluon GPD, the NN is constrained by the available lattice-QCD determinations of the pion gluon GFF. The corresponding total loss function is
\begin{equation}
\mathcal{L}_{\rm total}
=
\chi^2_{A_g}
+
\mathcal{L}_{\rm reg},
\end{equation}
where
\begin{equation}
\chi^2_{A_g}
=
\sum_i
\left[
\frac{
A_g^{\rm model}(t_i)
-
A_g^{\rm LQCD}(t_i)
}{
\sigma_i
}
\right]^2,
\end{equation}
with $A_g^{\rm model}(t_i)$ denoting the NN prediction obtained from
\begin{equation}
A_g(t)
=
\int_0^1
dx\,
H^g(x,t),
\end{equation}
and $A_g^{\rm LQCD}(t_i)$ represents the corresponding lattice-QCD result with uncertainty $\sigma_i$.

Because the gluon PDFs exhibit substantially larger uncertainties, particularly in the small-$x$ region, additional regularization terms are included during the optimization. These terms penalize large amplitudes, rapid variations, and nonphysical oscillatory behavior in the extracted gluon GFF, thereby improving both the numerical stability and the generalization capability of the NN model.

The minimization of the total loss function is performed using the Adam optimizer \cite{Kingma2015}, as implemented in the \texttt{PyTorch} framework \cite{Paszke2019}. The initial learning rate is chosen as
\begin{equation}
\eta_{\max}
=
5\times10^{-4},
\end{equation}
for the valence-quark analysis, while
\begin{equation}
\eta_{\max}
=
2\times10^{-4},
\end{equation}
is employed for the gluon analysis. In both cases, $L_2$ weight-decay regularization is included during the optimization to mitigate overfitting \cite{Krogh1992}. The network weights are initialized using the Xavier uniform initialization scheme \cite{Glorot2010}, while all bias parameters are initialized to zero. In addition, dropout regularization with a dropout probability of $0.05$ is applied after each hidden layer to improve the generalization capability of the NN.

To further stabilize the optimization, gradient clipping with a maximum norm of unity is employed throughout the training procedure \cite{Pascanu2013}. The learning rate is updated using a cosine-annealing scheduler \cite{Loshchilov2016}
\begin{equation}
\eta_t
=
\eta_{\min}
+
\frac12
\left(
\eta_{\max}
-
\eta_{\min}
\right)
\left[
1+
\cos
\left(
\frac{t\pi}{T_{\max}}
\right)
\right],
\end{equation}
where $\eta_t$ denotes the learning rate at epoch $t$, while $\eta_{\max}$ and $\eta_{\min}=10^{-7}$ represent the maximum and minimum learning rates, respectively. Here, $T_{\max}$ denotes the total number of training epochs. This scheduling strategy provides a smooth decay of the learning rate, suppresses oscillatory convergence, and improves the stability of the optimization.

Unless otherwise stated, all NN models are trained for $10^{5}$ epochs. The uncertainties of the extracted GPDs are obtained by propagating the central LHAPDF replica together with its corresponding $1\sigma$ uncertainty band through the complete NN fitting procedure.
%
%
\section{Results and Discussions}
\label{sec:results}
Using the proposed NN parameterization together with the pion valence-quark and gluon PDFs from the JAM21 and xFitter analysis, we perform fits to the available experimental and lattice-QCD data. The resulting pion EMFF is shown in Fig.~\ref{fig:fpi_q2_4}, where $F_\pi(t)$ and $|F_\pi(t)|^2$ are displayed in the left and right panels, respectively, as functions of $|t|$ on a logarithmic scale over the range $0.01$--$10~\mathrm{GeV}^2$ at $\mu^2 = 4~\mathrm{GeV}^2$. The results are compared with the complete compilation of experimental and lattice-QCD data summarized in Table~\ref{tab:chi2_dataset}.

The two momentum-transfer regimes probe complementary aspects of the pion structure. At small values of $|t|$, the FF is sensitive to the long-distance charge distribution of the pion and is directly related to the pion charge radius through $\langle r_\pi^2\rangle = -6\, dF_\pi(t)/dt|_{t=0}$. At large momentum transfer, the FF probes the short-distance dynamics of the pion and, within perturbative QCD, is expected to approach the asymptotic hard-scattering behavior. The extracted FF exhibits the expected smooth and monotonic decrease from unity toward zero, reflecting the transition from the soft to the hard regime without any indication of oscillatory or nonphysical behavior that would signal overfitting. No additional normalization penalty is required for the valence-quark sector because the NN correction is constrained to satisfy $\mathrm{NN}(x,0)=1$ for both the quark and gluon GPDs. Consequently, the forward-limit relations, $H^q(x,0)=q_v^\pi(x)$ and $H^g(x,0)=xg(x)$, are satisfied exactly by construction. As a result, the sum-rule constraint $F_\pi(0)=1$ in Eq.~(\ref{eq:ff_sumrule}) is automatically fulfilled without introducing an additional normalization term in the loss function. Further, The data-to-theory ratios remain close to unity over the entire fitted kinematic range. 


The fitted Regge slope parameters are found to be $c = 0.681$ for the JAM21 analysis and $c = 0.655$ for the xFitter analysis. It should be emphasized that the Regge slope is scale dependent and therefore depends on the factorization scale $\mu^2$ rather than representing a universal constant. The resulting $\chi^2/N$ values for the JAM21 and xFitter-based analysis are summarized in Table~\ref{tab:chi2_dataset}. The overall $\chi^2/N$ values are $1.42$ and $1.40$ for the JAM21 and xFitter analysis, respectively, evaluated using a total of $176$ data points. Among the individual datasets, those from Refs.~\cite{Bebek:1977pe,Dally:1981ur,JeffersonLabFpi:2000nlc,JeffersonLabFpi:2007vir} exhibit $\chi^2/N \gtrsim 2$ for both PDF parameterizations. Comparable values have also been reported in Refs.~\cite{Goharipour:2025zsw,Puhan:2026pkz}, suggesting that these tensions originate primarily from the experimental datasets themselves—for example, from normalization uncertainties or process-dependent systematic effects—rather than from limitations of the present NN parameterization. The extracted pion charge radius is $0.668~\mathrm{fm}$ for the JAM21 analysis and $0.667~\mathrm{fm}$ for the xFitter analysis. Both values are in good agreement with the Particle Data Group world-average value of $0.659~\mathrm{fm}$ reported in Ref.~\cite{ParticleDataGroup:2022pth}.

In Fig.\ref{fig:Ag_fit}, results are shown for the gluon sector, an overall $\chi^2/N = 0.84$ is obtained for the combined lattice-QCD data sets. To reproduce the gluon GFF, $A_g(t)$, normalization factors of $1.46$ and $2.07$ are required for the JAM21 and xFitter-based analysis, respectively. At $t=0$, the extracted gluon momentum fraction, $\langle x g(x)\rangle = A_g(0)$, is found to be $0.37$ for JAM21 and $0.26$ for xFitter, whereas the available lattice-QCD calculations predict $A_g(0)\approx 0.55$--$0.60$.

This discrepancy in the overall normalization motivates the introduction of a rescaling factor in the gluon-sector fit, which has been incorporated into all gluon GPDs results presented in this work. Physically, the mismatch likely originates from the limited experimental constraints on the small-$x$ gluon distribution in current global PDF analysis. Both the JAM21 and xFitter parameterizations rely on comparatively sparse data to determine the gluon PDF at low scales before DGLAP evolution to $\mu^2=4~\mathrm{GeV}^2$. Additional contributions may arise from systematic uncertainties in the lattice-QCD determination of $A_g(t)$, including nonperturbative renormalization matching and residual excited-state contamination in the extraction of the relevant three-point correlation functions. Understanding the origin of this normalization difference between PDF-based extractions and lattice-QCD calculations remains an important open problem that warrants further investigation from both the global-analysis and lattice-QCD perspectives. Despite this normalization difference, the JAM21 and xFitter based analysis yield remarkably similar gluon GPDs after the rescaling is applied, indicating that the NN component, $\mathrm{NN}(x,t)$, successfully captures the residual $x$- and $t$-dependence largely independently of the particular collinear gluon PDF used as input.

The physical reliability of the extracted GPDs is further supported by the numerical stability of the NN optimization, as illustrated in Figs.~\ref{fig:loss_components} and \ref{fig:training_loss}. The dominant loss components, $\chi^2_{F_\pi}$ and $\chi^2_{|F_\pi|^2}$, decrease by more than an order of magnitude during the first few thousand training epochs and reach stable plateaus well before $10^{5}$ epochs, indicating rapid convergence to a well-defined minimum. The normalization penalty, $\chi^2_{F_\pi(0)}$, exhibits a brief initial increase before settling to values of $\mathcal{O}(1)$, reflecting the optimization required to simultaneously satisfy the charge-normalization constraint, $F_\pi(0)=1$, and reproduce the measured momentum-transfer dependence of the pion FF. Once the Regge parameter $c$ and the NN parameters converge, both constraints are satisfied simultaneously. Throughout the training, the regularization contribution, $\mathcal{L}_{\rm reg}$, remains small and nearly constant, demonstrating that it serves only to suppress unphysical oscillations without influencing the physical fit.

An important observation is that the JAM21 and xFitter based training histories shown in Fig.~\ref{fig:training_loss} remain almost perfectly superimposed throughout the entire optimization process, both for the total loss and for the reduced $\chi^2/N$. This demonstrates that the optimization consistently converges to statistically equivalent minima irrespective of the choice of input PDF. Consequently, the close agreement between the two analysis reflects the intrinsic structure of the loss function defined by Eq.~(\ref{eq:ff_sumrule}) and the experimental constraints, rather than a dependence on the random initialization or a particular optimization trajectory.

The overall values of $\chi^2/N\approx1.4$ reported in Table~\ref{tab:chi2_dataset}, together with the smooth and monotonic evolution of the individual loss components shown in Fig.~\ref{fig:loss_components}, indicate that the NN architecture is well matched to the size and precision of the available dataset. No evidence of overfitting is observed, which would manifest itself through unstable values of the fitted Regge parameter or significant run-to-run variations in the extracted GPDs. Likewise, no indication of underfitting is present, as evidenced by the absence of systematic deviations in the data-to-theory ratios shown in Fig.~\ref{fig:fpi_q2_4}.

The observed numerical stability results from the combination of the physics-informed multiplicative parameterization,
$H^q(x,t)=q_v^\pi(x)\,e^{ct}\,\mathrm{NN}(x,t)$ for the valence-quark
and $H^g(x,t)=xg(x)\,e^{c_g t}\,\mathrm{NN}(x,t)$ for the gluon,
which requires the NN to learn only the residual corrections beyond the known PDF and Regge behavior, together with the optimization strategy described in Sec.~\ref{nnfitting}. Specifically, the exponential output activation guarantees positivity, while Xavier initialization, dropout regularization, gradient clipping, and cosine-annealing learning-rate scheduling collectively improve the stability and convergence of the training procedure.

The gluon-sector training history exhibits the same qualitative behavior as that observed for the valence-quark analysis, as given in the Appendix. Although the gluon fit begins with a substantially larger initial loss owing to the comparatively larger uncertainties of the input gluon PDFs, both the JAM21- and xFitter-based optimizations converge rapidly during the first few thousand epochs and remain nearly identical throughout the remainder of the training. This demonstrates that the stability of the proposed NN framework is not limited to the quark sector but also extends to the considerably more challenging gluon sector. Overall, these results demonstrate that the extracted pion quark and gluon GPDs constitute reproducible, numerically stable, and physically constrained solutions rather than artifacts of a particular optimization path.

In Fig.~\ref{fig:gpd_q2_4}, we compare the extracted quark GPDs obtained using the JAM21 and xFitter pion PDF parameterizations at the reference scale $\mu^2=4~\mathrm{GeV}^2$. The left panel shows the momentum-weighted quark GPD, $xH^q(x,t)$, as a function of the longitudinal momentum fraction $x$ for fixed values of $-t=0.1$, $0.5$, $1.0$, $2.0$, and $5.0~\mathrm{GeV}^2$. For all momentum transfers, the distribution vanishes in the limits $x\rightarrow0$ and $x\rightarrow1$ and reaches a maximum at intermediate momentum fractions, $x\approx0.3$--$0.5$, reflecting the characteristic shape of the pion valence-quark PDF in the forward limit. As $|t|$ increases, the overall magnitude of the GPD decreases monotonically over the entire $x$ range, while the position of the peak remains nearly unchanged. This behavior is consistent with the adopted parameterization, in which the forward PDF is modulated by a momentum-transfer-dependent profile. Compared with the phenomenological results of Ref.~\cite{Goharipour:2025zsw}, our extracted GPDs are systematically smaller over most of the $x$ range. This difference illustrates the sensitivity of the extracted GPDs to the choice of profile function, demonstrating that different profile-function parameterizations can produce appreciably different momentum-transfer dependences even when based on the same forward PDFs.

The right panel of Fig.~\ref{fig:gpd_q2_4} presents $H^q(x,t)$ as a function of $|t|$ for fixed values of $x=0.2$, $0.3$, $0.5$, and $0.8$. For all momentum fractions, the distributions exhibit dependence on $|t|$ at small momentum transfer before decreasing smoothly as $|t|$ increases. The ordering $H^q(x=0.2)>H^q(x=0.3)>H^q(x=0.5)>H^q(x=0.8)$ is preserved over the entire momentum-transfer range, directly reflecting the hierarchy already present in the forward valence-quark PDF. The results obtained using the JAM21 and xFitter PDFs are nearly indistinguishable, and the corresponding uncertainty bands remain narrow, indicating that the extracted quark GPDs are only weakly sensitive to the choice of the underlying valence-quark PDF.


The extracted gluon GPD, $H^g(x,t)$, is shown in Fig.~\ref{fig:gluongpd_q2_4}. Unlike the quark GPD, the gluon distribution exhibits a pronounced enhancement at small momentum fractions. The left panel displays the momentum-weighted gluon GPD, $xH^g(x,t)$, as a function of $x$ for fixed values of $t=-0.1$, $-0.5$, and $-1.0~\mathrm{GeV}^2$. It is clear from the figure that distribution shows maxima in the low-$x$ region, and then decreases toward $x\rightarrow1$, reflecting the well-known enhancement of gluons at small momentum fractions. In contrast to the quark sector, noticeable differences are observed between the JAM21 and xFitter based extractions, accompanied by substantially broader uncertainty bands. These differences primarily originate from the comparatively poor constraints on the pion gluon PDF, particularly in the low-$x$ region, and demonstrate the sensitivity of the extracted gluon GPD to the choice of the input gluon PDF.

The right panel of Fig.~\ref{fig:gluongpd_q2_4} shows the momentum-transfer dependence of $H^g(x,t)$ for fixed values of $x=0.001$, $0.01$, $0.1$, and $0.5$. For all momentum fractions, the gluon GPD decreases monotonically with increasing $|t|$, while its magnitude varies by nearly an order of magnitude across the selected values of $x$. The largest contributions arise at the smallest momentum fractions, indicating that the pion's gluon content is predominantly concentrated in the low-$x$ region. In contrast, the distribution at $x=0.5$ is strongly suppressed throughout the entire momentum-transfer range. Compared with the quark GPD, the gluon GPD exhibits significantly larger uncertainties, reflecting the limited experimental constraints on the pion gluon distribution. Consequently, the gluon sector constitutes the dominant source of uncertainty in the present extraction.

Figs. \ref{fig8_lineargluon} and \ref{gpd_H_vs_t_log} compare the extracted quark GPDs with the lattice-QCD calculations of Ref.~\cite{Ding:2024saz}. Fig.~\ref{fig8_lineargluon} presents $H^q(x,t)$ as a function of the longitudinal momentum fraction $x$ for several fixed values of $-t$. In the forward limit ($-t=0$), the phenomenological and lattice-QCD results exhibit excellent agreement over the entire $x$ range. As the momentum transfer increases, the GPD is progressively suppressed for both approaches. At larger values of $|t|$, the direct lattice-QCD results lie systematically above the NN extraction, whereas the renormalization-group-resummed (RGR) lattice results remain in much closer agreement with the phenomenological curves. A complementary comparison is shown in Fig.~\ref{gpd_H_vs_t_log}, where $H^q(x,t)$ is plotted as a function of $|t|$ for fixed values of $x=0.2$, $0.3$, $0.5$, and $0.8$. A similar trend is observed: the agreement between the phenomenological and lattice-QCD results is excellent at low and intermediate momentum fractions, while the discrepancy gradually increases toward larger $x$, becoming most pronounced at $x=0.8$. Overall, the extracted quark GPD is consistent with the RGR lattice calculation over most of the explored kinematic region, with the largest deviations confined to the large-$x$ and large-$|t|$ domain.

The transverse spatial structure of the pion is obtained by taking the Fourier transform\cite{Burkardt:2000za,Burkardt:2002hr} of the extracted quark and gluon GPDs into impact-parameter space. The resulting impact-parameter-dependent quark distribution, $2\pi b\,H^q(x,\mathbf{b})$, is shown in Fig.~\ref{fig:Hq_density_vs_b}. As the longitudinal momentum fraction increases from $x=0.2$ to $0.8$, the distribution becomes progressively narrower and more localized around the transverse center of the pion, while its overall magnitude decreases. The JAM21 and xFitter based extractions are nearly indistinguishable over the entire kinematic range, indicating that the transverse quark density is only weakly affected by the choice of the input valence-quark PDF.

A direct comparison of the quark and gluon impact-parameter distributions at $x=0.5$, shown in Fig.~\ref{fig:Hg_density_vs_b}, reveals that the quark density is substantially larger than the gluon density, while the latter is accompanied by considerably broader uncertainty bands. This reflects the dominance of valence quarks at large momentum fractions together with the comparatively weaker experimental constraints on the pion gluon distribution.

The corresponding two-dimensional transverse density distributions provide a more intuitive visualization of the pion structure. For both the JAM21- and xFitter-based extractions (Figs.~\ref{fig8_contour_JAM21} and \ref{fig_contour_xfitter}), the quark density becomes increasingly concentrated near the transverse origin as $x$ increases, demonstrating the expected correlation between longitudinal momentum fraction and transverse localization. In contrast, the gluon density distributions (Figs.~\ref{fig:contour_JAM21} and \ref{fig:contour_xFitter}) are strongly concentrated in the small-$x$ region. The gluon density reaches its maximum at $x=0.001$, decreases rapidly with increasing momentum fraction, and becomes strongly suppressed by $x=0.5$, consistent with the predominance of gluons at small $x$. Although the overall spatial distributions obtained using the JAM21 and xFitter inputs are qualitatively similar, the gluon sector exhibits a noticeably stronger dependence on the choice of the input PDF because of the larger uncertainties associated with the underlying gluon distributions. Overall, the impact-parameter analysis demonstrates a clear correlation between the longitudinal momentum fraction and the transverse spatial structure of the pion, with valence quarks dominating the large-$x$ region and gluons predominantly populating the small-$x$ regime.

\section{Conclusion}
\label{conclu}
In this work, we have developed a neural-network (NN) framework for the extraction of the pion unpolarized generalized parton distributions (GPDs) of both quarks and gluons. The quark GPDs are constrained using the available experimental and lattice quantum chromodynamics (QCD) data for the pion electromagnetic form factor together with the JAM21 and xFitter pion parton distribution functions (PDFs), while the gluon GPDs are extracted using lattice-QCD calculations of the pion gluon gravitational form factor.

The proposed framework combines the known forward-limit behavior of the pion PDFs with a trainable NN component, allowing the GPDs to be determined in a flexible and largely model-independent manner while preserving the fundamental theoretical constraints. Compared with conventional Regge-inspired parameterizations based on fixed functional forms, the NN approach provides greater flexibility in describing the nonperturbative structure of the pion without introducing a large number of phenomenological parameters.

The extracted Regge slope parameters are found to be smaller than unity for both the JAM21- and xFitter-based analysis, consistent with the expected momentum-transfer dependence of the pion GPDs. The resulting fits describe the available data well, yielding overall reduced chi-square values of approximately $\chi^2/N \approx 1.4$ for the quark sector and $\chi^2/N \approx 0.84$ for the gluon sector. Furthermore, the stable convergence of the individual loss components and the total loss function demonstrates the numerical robustness and reliability of the NN optimization.

The extracted quark and gluon GPDs exhibit the expected physical behavior over the explored kinematic region. The quark GPDs are concentrated at intermediate and large momentum fractions, whereas the gluon GPDs are dominated by the small-$x$ region and carry substantially larger uncertainties, reflecting the current limitations of the available gluon PDF constraints. The corresponding impact-parameter distributions reveal the transverse spatial structure of the pion, illustrating the increasing localization of partons with increasing longitudinal momentum fraction. Overall, the JAM21- and xFitter-based analysis lead to remarkably consistent results, demonstrating that the extracted GPDs are robust against the choice of the underlying collinear PDF parameterization.

%

To the best of our knowledge, this work represents the first application of a NN framework to the extraction of pion GPDs. The proposed methodology provides a promising foundation for future global analysis of hadron tomography and can be naturally extended to the nucleon, where the simultaneous extraction of the full set of quark and gluon GPDs from experimental and lattice-QCD data presents an important next step.

%
%
\section*{Acknowledgments}
The authors would like to thank Qi Shi and Heng-Tong Ding for kindly providing the lattice-QCD data for the valence-quark GPDs used in this work.


\bibliographystyle{apsrev}  
\bibliography{ref} 
\newpage
\section{Appendix}
\label{aper}
The valence-quark and gluon GPD networks employ different output activation functions to ensure stable training and physically meaningful behavior. For the valence-quark sector, the final activation function is given by
\begin{equation}
\mathrm{NN}_q(x,t)=\exp\!\left(\mathrm{NN}(x,t)\right),
\end{equation}
which guarantees a positive multiplicative correction to the forward PDF. For the gluon sector, the final activation function is chosen as
\begin{equation}
\mathrm{NN}_g(x,t)=
\exp\!\left[
\tanh\!\left(\mathrm{NN}(x,t)\right)
\frac{|t|}{1+|t|}
\right],
\end{equation}
where the bounded $\tanh$ activation suppresses excessively large corrections during training, while the factor $|t|/(1+|t|)$ ensures $\mathrm{NN}_g(x,0)=1$, thereby preserving the forward-limit constraint. The use of different output activation functions is a modeling choice adopted to improve the numerical stability and convergence of the independent NN parameterizations.

\begin{figure}[h]
    \centering
    \includegraphics[width=0.48\textwidth]{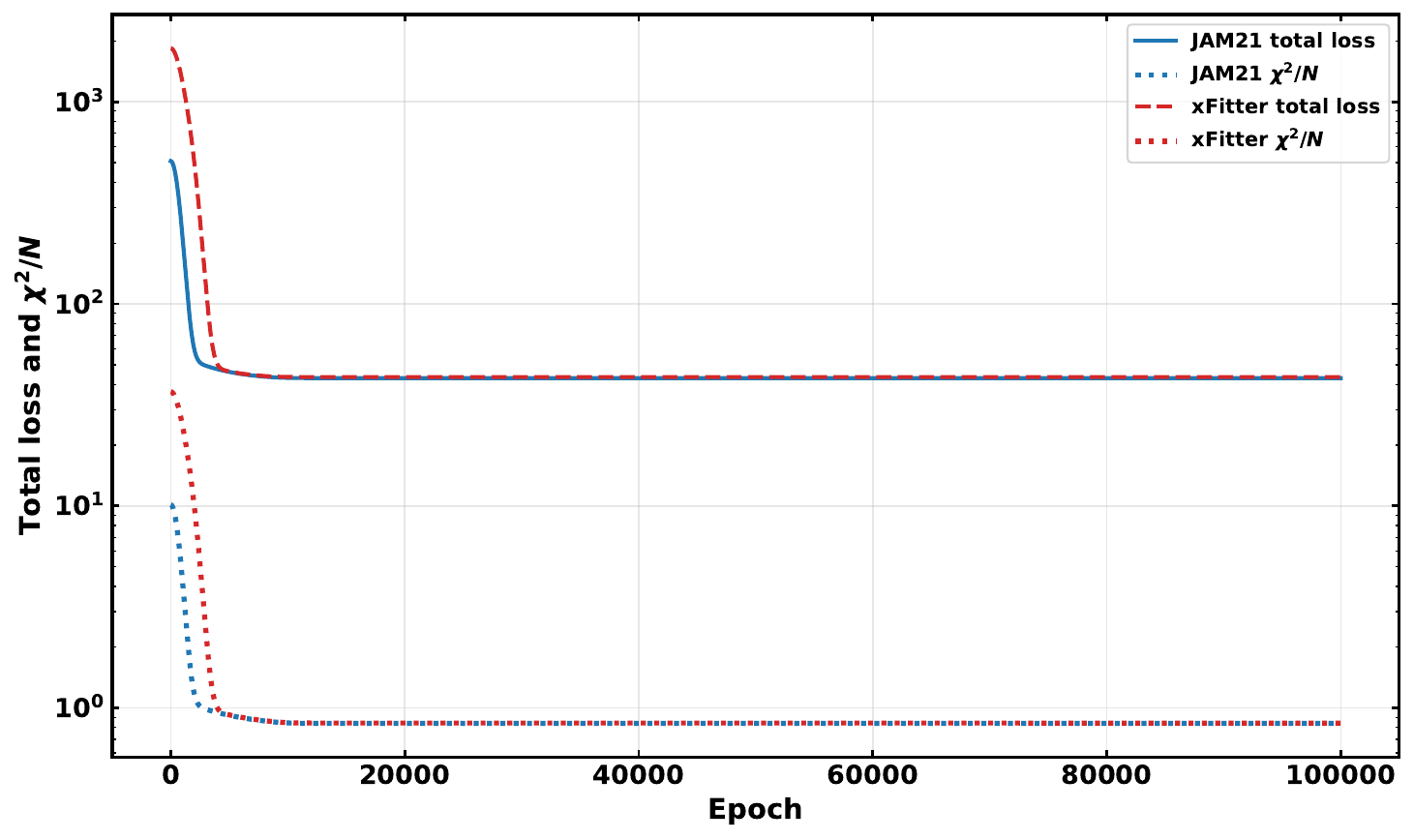}
    \caption*{
    Training history of the NN for the extraction of the pion gluon GPD at the scale $\mu^2 = 4~\mathrm{GeV}^2$. The figure shows the evolution of the total loss function and the corresponding normalized $\chi^2/N$ during training. The smooth decrease of both quantities demonstrates the stable convergence of the NN optimization.}
\end{figure}

\end{document}